\documentclass[11pt,a4paper]{article}
\usepackage{jheppubkim}
\usepackage{rotating} 
\usepackage{graphicx,epsfig}
\usepackage{amsmath}
\usepackage {amssymb}
\usepackage{subfigure}

\usepackage{relsize}

\newcommand{\be}{\begin{equation}}
\newcommand{\ee}{\end{equation}}
\newcommand{\bea}{\begin{eqnarray}}
\newcommand{\eea}{\end{eqnarray}}
\newcommand{\beaa}{\begin{eqnarray*}}
\newcommand{\eeaa}{\end{eqnarray*}}

\begin{document}

\title{Superbounce and Loop Quantum Ekpyrotic Cosmologies from Modified Gravity: $F(R)$, 
$F(G)$ and $F(T)$ Theories}

\author[a,b,c,d]{S.D. Odintsov}

\author[e,c,d]{V.K. Oikonomou}

\author[f,g]{Emmanuel N. Saridakis}

\affiliation[a]{Institut de Ciencies de l'Espai (IEEC-CSIC), Campus UAB,
Torre C5-Par-2a pl, E-08193 Bellaterra, Barcelona, Spain}

\affiliation[b]{InstituciŽo Catalana de Recerca i Estudis Avanžcats (ICREA), Barcelona, 
Spain}

 \affiliation[c]{National Research Tomsk State University, Tomsk, 634050}

\affiliation[d]{Tomsk State Pedagogical University,
Tomsk, 634061 Russia}

\affiliation[e]{Department of Theoretical Physics, Aristotle University of Thessaloniki,
54124 Thessaloniki, Greece}

\affiliation[f]{Physics Division, National Technical University of Athens,
15780 Zografou Campus,  Athens, Greece}

\affiliation[g]{Instituto de F\'{\i}sica, Pontificia
Universidad de Cat\'olica de Valpara\'{\i}so, Casilla 4950,
Valpara\'{\i}so, Chile}

\emailAdd{odintsov@ieec.uab.es}
\emailAdd{v.k.oikonomou1979@gmail.com}
\emailAdd{Emmanuel$_-$Saridakis@baylor.edu}

\abstract{We investigate the realization of two bouncing paradigms, namely of the 
superbounce and the loop quantum cosmological ekpyrosis, in the framework of various 
modified gravities. In particular, we focus on the $F(R)$, $F(G)$ and $F(T)$ 
gravities, and we reconstruct their specific subclasses which lead to such 
universe evolutions. These subclasses constitute from power laws, polynomials, or 
hypergeometric ansatzes, which can be approximated by power laws. The qualitative 
similarity of the different effective gravities which realize the above two bouncing 
cosmologies, indicates that a universality might be lying behind the bounce. Finally, 
performing a linear perturbation analysis, we show that the obtained solutions are 
conditionally or fully stable.}


\maketitle

\newpage
\section{Introduction}
\label{Introduction}

According to many theoretical arguments and observational indications, the universe 
exhibited an early accelerating phase called inflation 
\cite{Guth:1980zm,Linde:1981mu}. The inflationary paradigm, amongst 
others, is very efficient in solving the flatness, horizon and monopole problem, and 
offers a consistent mechanism for the generation of the primordial fluctuations or 
primordial 
gravitational waves \cite{Starobinsky:1982ee,Linde:1983gd} (see also 
\cite{mukhanov,GorbunovRubakov}).  
Hence, current observational research \cite{bicep,planck} is focused on revealing the 
Universe's evolution during all eras and theoretical research is focused on embedding all 
the striking new results in a consistent theoretical framework, and if possible under the 
same theoretical framework.

One alternative to the standard inflation description of the early acceleration is 
provided by bouncing cosmologies \cite{Novello:2008ra,Brandenberger:1993ef}, in which one 
avoids the appearance of the initial singularity. Some of the  
most appealing representative theories in this paradigm, are the ones that use  scalar 
fields 
\cite{Novello:2008ra,Brandenberger:1993ef,bounce3,bounce3b,bounce4,bounce5,Lehners:2008vx,
Qiu:2013eoa,Cai:2014xxa,Khoury:2001bz,Khoury:2001zk,Khoury:2001wf,Erickson:2003zm, 
Lehners:2013cka}, the ones that use various models of modified gravity 
\cite{Veneziano:1991ek, Brustein:1997cv,
Kehagias:1999vr, Shtanov:2002mb, Saridakis:2007cf,Creminelli:2007aq,
Cai:2010zma, HLbounce,Cai:2009in,
Bojowald:2001xe,Martin:2003sf,Saridakis:2010mf,Cai:2012ag}, and the ones that use the 
Loop 
Quantum Cosmology 
\cite{Ashtekar:2007tv,Ashtekar:2011ni,Bojowald:2008ik,Cailleteau:2012fy,Haro:2014wha, 
deHaro:2014kxa,Amoros:2014tha,Cai:2014zga, WilsonEwing:2012pu} (LQC) theoretical 
framework. In most of these scenarios, the bounce realization relies on the existence of 
matter fields, with an equation of state of suitable form in order to achieve the 
necessary violation of  the null energy condition 
\cite{Novello:2008ra,Brandenberger:1993ef,bounce4,bounce5}. 
Bouncing cosmologies may have primordial instabilities, which make the contracting phase 
a rather problematic era \cite{bounce3,bounce3b}, however one can solve this problem in 
many ways, as for instance in the framework of ekpyrotic scenario \cite{bounce5,bounce4}. 
For relevant works on bouncing cosmologies see \cite{bouncepiao,bouncepiao1,bounceref1} 
and for a study on the non-existence of open or flat bouncing cosmologies in Einstein 
gravity, see \cite{bounceref2}. Finally, 
note that there might be observable signatures of the bouncing phase 
\cite{kinezosvergados,oikonomouvergados,oikonomouvergados2}.  

In the present work we are interested in investigating the superbounce and the LQC 
ekpyrotic scenarios in the framework of modified gravity 
\cite{Nojiri:2006ri,Capozziellbook,Capozziello:2011et}. Since modified gravity is one of 
the two main directions in which one can achieve also the late-time acceleration (with 
the 
other one being dark energy), the above incorporation of bouncing solutions in modified 
gravity becomes more important under a unified picture, namely to be able to obtain the 
bouncing-initial phase of the universe as well as its late-time acceleration, 
simultaneously. In particular, we will investigate the above bouncing behaviors in the 
context of 
$F(R)$ 
\cite{Capozziello:2002rd,Carroll:2003wy,Capozziello:2003gx,Sotiriou:2006qn,
Capozziello:2006dj,Faraoni:2007yn,Hu:2007nk,Appleby:2007vb, 
Nojiri:2007as,Dunsby:2010wg,importantpapers3,recontechniques,
recontechniques1,recontechniques1b,recon3,sergeinojirimodel,
sergeibabmba} (for recent reviews see \cite{DeFelice:2010aj,Nojiri:2010wj}), $F(G)$  
\cite{Boulware:1985wk,Wheeler:1985nh,Antoniadis:1993jc,Kanti:1998jd,Nojiri:2005vv,
Nojiri:2005jg,Cognola:2006eg,Davis:2007ta, Eynard:2007nq,Jawad:2013wla,fg4,fg6}
and $F(T)$  
\cite{Linder:2010py,Chen:2010va,Dent:2011zz,Bamba:2010wb,Zhang:2011qp,Sharif001,
Capozziello006,Geng:2011aj,Bohmer:2011si,Gonzalez:2011dr,Karami:2012fu,Bamba:2012vg,
Iorio:2012cm,
Rodrigues:2012qua,Capozziello:2012zj,Chattopadhyay:2012eu,Izumi:2012qj,Li:2013xea,
Ong:2013qja,
Otalora:2013tba,
Nashed:2013bfa,Kofinas:2014owa,Harko:2014sja,Hanafy:2014bsa,
Junior:2015bva,Ruggiero:2015oka} 
theories. The techniques that we shall use in order to provide a modified gravity 
description of the bouncing cosmologies are quite renowned 
\cite{recontechniques,importantpapers3,oikonomoubounce} (see also 
\cite{recontechniques1,recontechniques1b,recon3}). For related works on bounce cosmology 
reconstruction from $F(R)$ theories see \cite{mbs,mbm}, and for $F(R)$ ekpyrotic 
cosmology 
see \cite{sekpd}. Finally, for completeness, we investigate whether the obtained 
solutions are stable under linear perturbations.

The manuscript is organized as follows: In section \ref{FRrecon}, using very well known 
reconstruction techniques, we investigate the $F(R)$ modified gravity, and in particular 
we extract the $F(R)$ forms that give rise superbounce and LQC ekpyrotic evolution. For 
the superbounce case we provide a complete analytic solution in closed form, while for 
the case of LQC ekpyrosis we extract the
large and small curvature limits. In Section  \ref{FGrecon} we perform the same 
analysis for the case of $F(G)$ gravity, and similarly, in 
Section \ref{FTrecon} for the $F(T)$ 
gravitational paradigm. In Section \ref{Stability} we investigate the stability of the 
obtained   $F(R)$, $F(G)$ and $F(T)$ solutions. With regards to the $F(R)$ and 
$F(G)$, the stability analysis is focused on examining the dynamical system formed by the
resulting FRW equations, under linear perturbations. Additionally, the resulting 
conditions that ensure stability are thoroughly investigated. Finally, Section 
\ref{Conclusions} is devoted to the conclusions.

\section{Superbounce and loop quantum ekpyrotic cosmology from $F(R)$ gravity}
\label{FRrecon}

In this Section we will reconstruct the suitable $F(R)$ form that can lead to the 
superbounce and loop quantum ekpyrotic cosmology realizations.
Before investigating the reconstruction procedure, we briefly review in the following 
subsection the basic features of $F(R)$ gravity in the metric formalism and we 
describe the geometric background we shall use in its cosmological application. For a 
detailed account on these issues, see \cite{DeFelice:2010aj,Nojiri:2010wj} and references 
therein.

\subsection{$F(R)$ gravity and cosmology}

Let us review briefly the $F(R)$ gravitational modification and its cosmological 
application. Throughout the article we assume that the spacetime manifold is 
pseudo-Riemannian, locally described by a Lorentz metric. Moreover, we consider a 
torsion-less, symmetric, and metric compatible affine connection, namely the Levi-Civita 
one. The Christoffel symbols write as
\begin{equation}\label{christofell}
\Gamma_{\mu \nu }^k=\frac{1}{2}g^{k\lambda }(\partial_{\mu }g_{\lambda \nu}+\partial_{\nu
}g_{\lambda \mu}-\partial_{\lambda }g_{\mu \nu})
\end{equation} 
while the Ricci scalar becomes
\begin{equation}\label{ricciscalar}
R=g^{\mu \nu }(\partial_{\lambda }\Gamma_{\mu \nu }^{\lambda}-\partial_{\nu }\Gamma_{\mu 
\rho
}^{\rho}-\Gamma_{\sigma \nu }^{\sigma}\Gamma_{\mu \lambda }^{\sigma}+\Gamma_{\mu \rho 
}^{\rho}g^{\mu
\nu}\Gamma_{\mu \nu }^{\sigma}).
\end{equation}
The four dimensional action that describes $F(R)$ theories
then writes as
\begin{equation}\label{action}
\mathcal{S}=\frac{1}{2\kappa^2}\int \mathrm{d}^4x\sqrt{-g}F(R)+S_m,
\end{equation}
with $\kappa^2=8\pi G$ is the gravitational constant, and  where we have also considered 
the action of the matter sector $S_m$. In the context 
of the 
metric formalism, by varying action (\ref{action}) with respect to the metric tensor 
$g_{\mu \nu}$ 
we obtain the equations of motion:
\begin{equation}\label{eqnmotion}
F'(R)R_{\mu \nu}(g)-\frac{1}{2}F(R)g_{\mu \nu}-\nabla_{\mu}\nabla_{\nu}F'(R)+g_{\mu 
\nu}\square
F'(R)=\kappa^2T_{\mu \nu}^m.
\end{equation} 
In the above expression and throughout the paper, the prime is assumed to denote 
differentiation with respect to the corresponding argument (hence in the present case we 
have $F'(R)=\partial F(R)/\partial R$). In addition, $T_{\mu \nu}^m$ stands for the 
matter energy-momentum tensor arising from $S_m$.

In order to investigate the cosmological implications of the above gravitational theory, 
we consider as usual a flat Friedmann-Lemaitre-Robertson-Walker (FRW) metric, with line 
element
\begin{equation}\label{metricformfrwhjkh}
\mathrm{d}s^2=-\mathrm{d}t^2+a^2(t)\sum_i\mathrm{d}x_i^2.
\end{equation}
In this case, the Ricci scalar is calculated to be
\begin{equation}\label{ricciscal}
R=6(2H^2+\dot{H}),
\end{equation}
where $H=\dot{a}/a$ stands for the Hubble parameter and  dots indicate differentiation 
with respect to the cosmic time $t$. 
Hence, the field equations (\ref{eqnmotion}), in the case of FRW geometry give 
rise to the Friedmann equation, namely
\begin{equation}\label{frwf1}
-18\left [ 4H(t)^2\dot{H}(t)+H(t)\ddot{H}(t)\right ]F''(R)+3\left [H^2(t)+\dot{H}(t) 
\right ]F'(R)-\frac{F(R)}{2}+\kappa^2\rho_{tot}=0,
\end{equation}
with $\rho_{tot}$ the total  energy density of the matter fields, and where the Ricci 
scalar 
is given in (\ref{ricciscal}). 

\subsection{Superbounce Reconstruction from $F(R)$ Gravity}
\label{FRsuperbouncereconstr}

The superbounce cosmological scenario was studied in \cite{bounce4}, both in a 
supergravity and non-supersymmetric framework, with the most appealing attribute of this 
scenario being the ekpyrotic contraction phase. The ekpyrotic contraction was firstly 
introduced and studied in \cite{bounce5}, along with a thorough study on cosmological 
perturbations. The purpose of this Section is to provide an $F(R)$ description of the 
superbounce cosmological scenario, and in order to achieve this we will make use of the 
reconstruction technique firstly developed in \cite{importantpapers3}. 

The superbounce scale factor is given by \cite{bounce4}
\begin{equation}\label{basicsol1}
a(t)\sim (-t+t_*)^{2/c^2},
\end{equation}
with $t_*$ being the big crunch time, and $c$ a parameter constrained to $c>\sqrt{6}$ 
\cite{bounce4}. Thus, the corresponding Hubble parameter is
\begin{equation}\label{hubble1}
H(t)=\frac{2}{c^2(t-t_*)}.
\end{equation}
The reconstruction technique developed in \cite{importantpapers3} is an exact method 
based on the introduction of a new variable in place of the cosmological time $t$, namely 
the e-folding number $N$, related to the scale factor $a(t)$ through
\begin{equation}\label{efoldpoar}
e^{-N}=\frac{a_0}{a}.
\end{equation} 
Using the new variable $N$, we can re-express the first FRW equation (\ref{frwf1}) in 
terms of the e-fold parameter $N$ as
\begin{eqnarray}
\label{newfrw1}
&& 
\!\!\!\!\!\!\!\!\!\!\!\!\!\!\!\!\!\!\!\!\!\!\!\!\! 
-18\left [ 4H^3(N)H'(N)+H^2(N)(H')^2+H^3(N)H''(N) \right ]F''(R)\notag
\\  
&& \ \ \ \ \ \ \ \ \ +3\left [H^2(N)+H(N)H'(N) \right]F'(R)-\frac{F(R)}{2}+\kappa^2\rho=0.
\end{eqnarray}
We mention that in relation (\ref{newfrw1}), and in the rest of this subsection, the 
Hubble parameter is assumed to be a function of the e-folds $N$ and in addition any 
derivative appearing in (\ref{newfrw1}) is defined with respect to $N$.

We now introduce the function $G(N)=H^2(N)$, and therefore we have that
\begin{equation}\label{riccinrelat}
R=3G'(N)+12G(N).
\end{equation}
This relation is very convenient, since it allows us to determine $N(R)$. In particular, 
the Hubble parameter (\ref{hubble1}) can be written in terms of the scale factor as
\begin{equation}\label{hpscf}
H=\frac{2}{c^2}a^{-\frac{c^2}{2}},
\end{equation}
and thus eliminating $a$ in favor of $N$ using  (\ref{efoldpoar})  we obtain
\begin{equation}\label{gnfunction}
G(N)=A e^{-c^2N}
\end{equation}
where we have set $A=\frac{4}{c^4}a_0^{-c^2}$. Hence, by combining relations 
(\ref{riccinrelat}) and (\ref{gnfunction}), we can acquire the e-fold parameter $N$ as a 
function of $R$ as
\begin{equation}\label{efoldr}
N=-\frac{1}{c^2}\ln \left[\frac{R}{3A(4-c^2)}\right].
\end{equation}
  Inserting these in the Friedmann equation  (\ref{newfrw1}),
we can re-express it as
 \begin{eqnarray}
\label{newfrw1modfrom}
&& 
\!\!\!\!\!\!\!\!\!\!\!\!\!\!\!\!\! 
-9G(N(R))\left[ 4G'(N(R))+G''(N(R)) 
\right]F''(R) \notag
\\ && \ \ \ \ \ \ \ \ \ 
+\left[3G(N)+\frac{3}{2}G'(N(R)) 
\right]F'(R)-\frac{F(R)}{2}+\kappa^2\rho_{tot}=0,
\end{eqnarray}
with $G'(N)=\mathrm{d}G(N)/\mathrm{d}N$ and $G''(N)=\mathrm{d}^2G(N)/\mathrm{d}N^2$. 
The last step is to express the total matter energy density $\rho_{tot}$ in terms of 
$N(R)$. 
Assuming the various matter components $\rho_i$ to be independently conserved, i.e. 
$\dot{\rho}_i+3H(1+w_i)\rho_i=0$, with $w_i$ their corresponding equation-of-state 
parameters, we obtain
\begin{equation}\label{mattenrgydens}
\rho_{tot} =\sum_i\rho_{i0}a_0^{-3(1+w_i)}e^{-3N(R)(1+w_i)}.
\end{equation}
Hence, inserting all the above in the Friedmann equation (\ref{newfrw1modfrom}), we 
obtain a differential equation for $F(R)$, namely
\begin{align}
\label{bigdiffgeneral1}
&a_1 R^2\frac{\mathrm{d}^2F(R)}{\mathrm{d}R^2}
+a_2R\frac{\mathrm{d}F(R)}{\mathrm{d}R}-\frac{F(R)}{2}+\sum_iS_{i}R^{
\frac{3(1+w_i)
}{c^2}}=0,
\end{align}
where we have set
\begin{eqnarray}
\label{apara1a2}
&&a_1=\frac{c^2}{4-c^2}\nonumber\\
&&a_2=\frac{2-c^2}{2(4-c^2)},
\end{eqnarray}
and
\begin{equation}\label{gfdgfdgf}
S_i=\frac{\kappa^2\rho_{i0}a_0^{-3(1+w_i)}}{[3A(4-c^2)]^{\frac{3(1+w_i)}{c^2}}} .
\end{equation}
 The  solution of (\ref{bigdiffgeneral1}) provides the exact $F(R)$ form that 
produces the superbounce evolution (\ref{basicsol1}).

Let us first consider pure $F(R)$ gravity, with no matter fluids present. In this case
the differential equation (\ref{bigdiffgeneral1}) becomes a homogeneous Euler 
second-order differential equation, with  solution 
\begin{equation}\label{frgenerlargetssss}
F(R)=c_1R^{\rho_1}+c_2R^{\rho_2},
\end{equation}
where $c_1,c_2$ are arbitrary parameters, and $\rho_1$ and $\rho_2$ are equal to 
\begin{eqnarray}\label{rho12}
&&\rho_1=\frac{-(a_2-a_1)+\sqrt{(a_2-a_1)^2+2a_1}}{2a_1}\nonumber\\
&&\rho_2=\frac{
-(a_2-a_1)-\sqrt{(a_
2-a_1)^2+2a_1}}{2a_1}.
\end{eqnarray}
Finally, it is worthy to examine if this $F(R)$ gravity can become Einstein gravity plus 
curvature terms under some parameter values. This indeed occurs when $c\gg 1$, 
in which case the $F(R)$ gravity takes the form, 
\begin{equation}\label{frgenerlargetsssslargec}
F(R)\simeq c_1R+c_2R^{-1/2}.
\end{equation}

We now come to the more physically interesting case where matter fields are present.
In this case, the general solution of  (\ref{bigdiffgeneral1}) is easily found to be 
\begin{eqnarray}
\label{newsolutionsnoneulerssss}
&&F(R)=\left 
[\frac{c_2\rho_1}{\rho_2}-\frac{c_1\rho_1}{\rho_2(\rho_2-\rho_1+1)}\right]R^{\rho_2+1}
+\sum_i \left[\frac{c_1S_i}{\rho_2(\delta_i+2+\rho_2-\rho_1)}\right] 
R^{\delta_i+2+\rho_2} 
\notag
\\ 
&& \ \ \ \ \ \ \ \ \
\  -\sum_iB_ic_2R^{\delta_i+\rho_2}+c_1R^{\rho_1}+c_2R^{\rho_2}
\end{eqnarray}
where $c_1,c_2$ arbitrary parameters, and where we have set
\begin{eqnarray}
\label{paramefgdd}
&&\delta_i=\frac{3(1+w_i)-2c^2}{c^2}-\rho_2+2\nonumber\\
&&B_i=\frac{S_i}{\rho_2\delta_i}.
\end{eqnarray}
It is interesting to note that for  $c\gg 1$ the above $F(R)$ gravity, responsible for 
the superbounce generalization in the present pf matter fields, becomes 
\begin{align}\label{newsolutionsnoneulersssslargec}
F(R)\simeq R+\alpha R^{2}+c_1R^{-1/2}+\Lambda,
\end{align}
where we have set $c_2=1$, $\alpha=\frac{c_1}{3}-2+2c_1\sum_i\frac{S_i}{3}$ and 
$\Lambda=-\sum_iA_
i$.

The $F(R)$ gravity (\ref{newsolutionsnoneulersssslargec}) can describe both early and 
late time acceleration, since at early times is approximately equal to $R+\alpha R^2$ 
(Starobinsky inflation) and at late times 
$R-R^{-1/2} +\Lambda$. For a detailed study of such $F(R)$ forms, and the 
phenomenological determination of the involved parameters, see   
\cite{sergeinojirimodel,sergeistarobinsky}).  

Let us close this subsection by a remark. In reference \cite{oikonomoubounce} a 
different reconstruction procedure was used in order to obtain, in the large and small 
curvature limits, the $F(R)$ gravity that produces the superbounce cosmological solution. 
The resulting $F(R)$ gravity in the large $R$ limit is quite similar to the one we 
obtained here, namely of the form $R+\alpha R^2$. This was expected to some extent, since 
the two reconstruction methods overlap at certain limits. This overlap however, is in 
some 
cases, 
model dependent and it is strongly affected by the particular form of the given Hubble 
parameter, and in particular from the limit of the Hubble parameter at large and 
small cosmological times. Later on we shall discuss on this issue in more detail.

\subsection{Loop quantum cosmology ekpyrotic scenario reconstruction from $F(R)$ gravity}
\label{FRLQCreconstr}

One quite appealing alternative to standard inflation is provided by the ekpyrotic 
scenario 
\cite{Khoury:2001bz,Khoury:2001zk,Khoury:2001wf,Erickson:2003zm,Lehners:2013cka}, in 
which scale invariant perturbations are generated before the Big-Bang phase. A recent 
refinement of the ekpyrotic scenario was presented in \cite{WilsonEwing:2012pu}, in which 
case LQC modifications to the original ekpyrotic 
scenario are taken into account 
\cite{Bojowald:2001xe,Ashtekar:2007tv,Ashtekar:2011ni,Bojowald:2008ik,Cailleteau:2012fy,
Haro:2014wha, deHaro:2014kxa,Amoros:2014tha,Cai:2014zga, WilsonEwing:2012pu}. In the 
context of LQC, the ekpyrotic scenario is realized by using the scalar field 
potential \cite{WilsonEwing:2012pu}
\begin{equation}
\label{ekpypot1}
V(\phi )=-\frac{V_0e^{\sqrt{16\pi G \rho}\phi}}{\left [1+\frac{3\rho V_0}{4\rho_c(1-3\rho 
)}e^{\sqrt{16\pi G \rho}\phi} \right ]^2},
\end{equation}
with the parameter $\rho$ taking values $0<\rho\ll 1$. As we shall explicitly show in the 
following Sections, the fact that $\rho\ll 1$ has particularly appealing consequences 
with respect to early and late time cosmological phenomenology, that are absent in 
other bouncing solutions, as for instance in the matter bounce scenario. Since this issue 
is very important, we shall thoroughly discuss it in the end of this Section. 


Before proceeding to the quantitative investigation of the reconstruction procedure, we 
need to make some important clarifications. The potential (\ref{ekpypot1}) is crucial for 
our analysis, since it guarantees the existence of the ekpyrotic contraction itself. 
However, this  potential was derived under the assumption that a phase of ekpyrosis 
actually exists in loop quantum cosmology. Nevertheless, as analytical and numerical 
investigations in the context of loop quantum cosmology indicate 
\cite{referee1,referee2}, 
for a Steinhardt-Turok type ekpyrotic potential, a phase with equation-of-state parameter 
$w\gg 1$ may not even exist, and consequently there may be no ekpyrosis. The reason for 
this is traced to the behavior of the moduli field's kinetic energy. Particularly,  in 
loop quantum cosmology during the contracting phase, and as the moduli field's kinetic 
energy approaches Planckian energies on the way towards the classical singularity, the 
Universe bounces. Hence, the Universe may never get a chance to reach the $w\gg 1$ 
regime, and thus it will bounce in the regime where $w\sim 1$ \cite{referee1,referee2}. 
Therefore, it is possible that for a large range of parameters in loop quantum
cosmology, for the Ekpyrotic potential of Steinhardt-Turok type, it is rather difficult to
obtain ekpyrosis. For an extensive and informative discussion and analysis of these 
issues the reader is referred to Refs. \cite{referee1,referee2}. 

Having these in mind, in the following we assume that an ekpyrotic phase, described by 
the 
potential (\ref{ekpypot1}), actually exists and thus in the LQC ekpyrotic scenario the 
corresponding scale factor and the Hubble parameter are equal to
\begin{equation}\label{scalf}
a(t)=\left ( a_0t^2+1 \right )^{\rho/2},{\,}{\,}{\,}H(t)=\frac{2a_0\rho t}{a_0t^2+1},
\end{equation}
with $\rho_c$ the critical density and $a_0=\frac{8\pi G\rho_c }{3\rho^2}$. It is the 
purpose of this subsection to find the $F(R)$ gravity that generates the cosmology 
described by relations (\ref{scalf}). As we shall demonstrate, we can obtain the 
cosmology (\ref{scalf}) without the need for a potential, hence modified gravity provides 
the theoretical framework which can describe ekpyrosis in the situation when the 
corresponding potential is questionable, or does not work completely successfully. 


Following closely the lines of the previous subsection, in the case at hand the Hubble 
parameter as a function of the scale factor is given by
\begin{equation}\label{hpscfss}
H^2=4\rho^2a_0\left (a^{\frac{6}{\rho}}-a^{\frac{4}{\rho}}\right),
\end{equation}
and by using (\ref{efoldpoar}) and also the function $G(N)=H^2(N)$, we obtain
\begin{equation}\label{gnfunctionss}
G(N)=A\left (\frac{4}{3}\rho_ce^{\frac{6N}{\rho}}-e^{\frac{4N}{\rho}} \right ),
\end{equation}
where for simplicity we have set the initial scale factor equal to one and also 
$A=4\rho^2a_0$. 
Thus, using relations (\ref{riccinrelat}) and (\ref{gnfunctionss}) we can express the 
e-fold 
number $N$ as a function of the Ricci scalar $R$ as
\begin{eqnarray}
&& \!\!\!\!\!\!\!\!\!\!\!\!\!\!\!\!\!\!\!\!\!\!\!\!\!\!\!
N=\frac{\rho}{2}\ln \Big\{\left\{6A^{\frac{5}{3}} (3+2 \rho ) \left[-32 A (1+\rho )^3
-9   R \rho  (3+2 \rho )^2+M_1\right]^{1/3}\right\}^{-1}  \notag 
\\ \ \ \ \ \ \ \ 
&&  \times   \Big\{-8 2^{2/3} A^2 
(1+\rho )^2+4 A^{\frac{5}{3}} (1+\rho ) 
  \left[-32 A (1+\rho )^3-9  R \rho  (3+2 \rho )^2+M_1\right]^{1/3}\notag
\\
&& \ \ \ \ \ 
-2^{1/3}A^{\frac{4}{3}} \left[-32 A (1+\rho )^3-9 R \rho  (3+2 \rho 
)^2+M_1\right]^{2/3}\Big\} 
 \Big\},
\label{efoldrss}
\end{eqnarray}
with $M_1=3 \sqrt{ R 
\rho  (3+2 \rho )^2 \left[64 A (1+\rho )^3+9 R \rho  (3+2 \rho )^2\right]}$.

By substituting expression (\ref{efoldrss}) into equation (\ref{newfrw1}), we acquire
\begin{align}\label{bigdiffgeneral1ss}
& 
A_1(\rho,R)\frac{\mathrm{d}^2F(R)}{\mathrm{d}R^2}+A_2(\rho,R)\frac{\mathrm{d}F(R)}{\mathrm
{d}R}-\frac{F(R)}{2}+\rho_m=0
\end{align}
with $\rho_m$ encompassing the contribution of all matter fluids to matter-energy 
density, and where the  coefficients $A_i(\rho,R)$, $i=1,2$
read
\begin{eqnarray}
&& \!\!\!\!\!\!\!
A_1(\rho,R)=- \left[432 A^4 \rho ^2 (3+2 \rho )^5 M_1^{10/3}\right]^{-1}\Big[ 
2^{11/3} A^2 (1+\rho )^2-4 A (1+\rho ) M_2^{1/3}
 +2^{1/3} M_2^{2/3}\Big]^4 
 \notag
\\  
 &&\ \ \ \ \ \ \ \  \ \ \ \,
 \times \Big\{128 A^4 
(1+\rho )
^4-12 A (1+\rho ) A^2M_1-3 A^2M_1 (2M_2)^{1/3}
\notag
\\  
 &&\ \ \ \ \ \ \ \  \ \ \ \ \ \ \, \
 +A^2  (2M_2)^{1/3}  \left[9 R \rho  (3+2 \rho )^2-8 
(1+\rho )^2 
 (2M_2)^{1/3}\right]
\notag
\\   
 &&\ \ \ \ \ \ \ \  \ \ \ \ \ \ \, \
+4 A^3 (1+\rho ) \left[9 R \rho  
(3+2 
\rho )^2
+8 (1+\rho )^2  (2M_2)^{1/3}\right]\Big\}
\notag
\\  
 &&\ \ \ \ \ \ \ \  \ \ \ \,
\times
\Big\{64 A^4 (1+\rho )^3 (7+4 
\rho )-6 A 
(7+4 \rho 
) A^2M_1  -3 A^2M_1  (2M_2)^{1/3}\nonumber\\
 &&\ \ \ \ \ \ \ \  \ \ \ \ \ \ \, \
+A^2  (2M_2)^{1/3}  
\left[9 R \rho  (3+2 \rho )^2-8 
(1+\rho )^2 
 (2M_2)^{1/3}\right]
\notag
\\  
 &&\ \ \ \ \ \ \ \  \ \ \ \ \ \ \, \
 +2 A^3  \left[9 R \rho  (3+2 \rho )^2 (7+4 \rho )+16 
(1+\rho )^3  (2M_2)^{1/3}\right]\Big{)}\Big\}
\label{coef1}
\end{eqnarray}
and 
\begin{eqnarray}
&&\!\!\!\!\!\!\!\!\!\!\!\!\!\! 
A_2(R,\rho)=  \left[72 A^2 \rho  (3+2 \rho )^3 M_2^{5/3}\right]^{-1}\Big[ 2^{11/3} A^2 
(1+\rho )^2-4 A (1+\rho ) M_2^{1/3}+2^{1/3} M_2^{2/3}
\Big]^2
\notag
\\  
&& \ \ \ \ \ \ \ \ \,
\times
\Big\{2A^3[12+\rho  (13+4 \rho )] [32 A (1+\rho )^3+ 9 R \rho  (3+2 \rho )^2
-3  M_1 ]
\notag
\\  
&& \ \ \ \ \ \ \ \ \ \ \ \ \
-3 (3+\rho ) A^2M_1 (2M_2)^{1/3}
+   32 A^3 (1+\rho )^3 (3+\rho )(2M_2)^{1/3}
\notag
\\ 
&& \ \ \ \ \ \ \ \ \ \ \ \ \
+A^2 (3+\rho ) (2M_2)^{1/3} [9 R \rho  (3+2 
\rho )^2
 -8 (1+\rho )^2 (2M_2)^{1/3}]
 \Big\}
\label{coef2}
\end{eqnarray}
 with
 $M_1=3 \sqrt{ R 
\rho  (3+2 \rho )^2 \left[64 A (1+\rho )^3+9 R \rho  (3+2 \rho )^2\right]}$ and
$M_2=3 A^2M_1-32 A^3 (1+\rho )^3-9 
A^2 R \rho  (3+2 \rho )^2$.

Solving explicitly the differential equation 
(\ref{bigdiffgeneral1ss}) is a challenging and rather formidable task, and thus in order 
to proceed we shall examine certain limits. Particularly, we shall study the large and 
small cosmological time limits of the scale factor (\ref{scalf}), and by applying the 
same 
reconstruction technique we shall investigate which $F(R)$ gravity generates 
corresponding limits. Note that the large and small time limits correspond to small and 
large curvature limits respectively.

\subsubsection{Large-time approximation of LQC ekpyrotic scenario}

Let us start with the large-time limit. In this case the scale factor and the Hubble 
parameter of (\ref{scalf}) become 
\begin{equation}\label{largetlimitofbsolu}
a(t)\approx a_0t^{\rho},{\,}{\,}{\,}H(t)\approx\frac{\rho}{t},
\end{equation}
and therefore the function $G(N)=H^2$ writes as
\begin{equation}\label{scalegn}
G(N)=\rho^2e^{-\frac{2N}{\rho}}.
\end{equation}
Inserting (\ref{scalegn}) into equation (\ref{riccinrelat}) and solving with respect to 
$R$ we obtain
\begin{equation}\label{nrlarget}
N=\frac{\rho}{3}\ln \left[\frac{6 \rho(2\rho-1)}{ R}\right].
\end{equation} 
Hence, substituting (\ref{nrlarget}) into equation (\ref{newfrw1modfrom}), we 
finally acquire the following differential equation:
 \begin{equation}
 \label{diffelargetapprox}
\left( \frac{1}{2\rho-1}\right)R^2F''(R) 
+\left(\frac{\rho-1}{4\rho-2}\right)RF'(R)-\frac{F(R)}{2}+\sum_i\rho_{0i}S_i 
R^{\alpha_i}=0,
\end{equation}
with
\begin{eqnarray}
&&S_i=\left[ 
6a_0\rho^2(2\rho-1) 
\right]^{-\frac{3(1+w_i)\rho}{2}}\nonumber\\
&&\alpha_i=\frac{3(1+w_i)\rho}{2}.\label{auxilparameters}
\end{eqnarray}

The differential equation (\ref{diffelargetapprox}) is the Euler non-homogeneous 
differential equation (\ref{bigdiffgeneral1}) which we came across in the previous 
subsection, thus in pure $F(R)$ gravity (disregarding all matter fluids), we get the 
following solution:
\begin{equation}\label{frgenerlarget}
F(R)=c_1R^{\rho_1}+c_2R^{\rho_2},
\end{equation}
with $c_1,c_2$ arbitrary parameters and with $\rho_1$ and $\rho_2$ given by
\begin{equation}\label{rho1rho2larget}
\rho_1=\frac{3-\rho-\sqrt{1+10\rho+\rho^2}}{4},{\,}{\,}{\,}\rho_2=\frac{3-\rho+\sqrt{
1+10\rho+\rho^2}}{4}.
\end{equation}
Since by construction in the context of the LQC ekpyrotic scenario it is required
$0<\rho\ll 1$, the solution (\ref{frgenerlarget}) is approximated to
\begin{equation}\label{frgenerlargetlimitbehvr}
F(R)=c_1R+c_2R^{1/2}.
\end{equation}

 By taking into account the presence of ordinary 
matter fields, and using the technique we employed in the previous subsection, the 
reconstructed $F(R)$ gravity which is solution of (\ref{diffelargetapprox}) 
reads
\begin{eqnarray}
&&\!\!\!\!\!\!\!\!\!\!\!\!\!\!\!\!
F(R)=c_1\sum_i 
\frac{S_i}{\rho_2(\alpha_i-\rho_1)}R^{\alpha_i}-\frac{c_1\rho_1}{\rho_2(\rho_2-\rho_1+1)
}R^{\rho_2+1}+c_1R^{\rho_1}\notag 
\\
&&\ \ \  -\sum_i 
\frac{S_ic_2}{\rho_2(\alpha_i-\rho_2)}R^{\alpha_i}+\frac{\rho_1c_2}{\rho_2}R^{\rho_
2+1}+c_3R^{\rho_2},
\label{newsolutionsnoneuler}
\end{eqnarray}
and therefore for $0<\rho\ll 1$ it becomes
\begin{equation}\label{newsolutionsnoneulersmallrholimit}
F(R)=c_1R+\frac{\rho_1}{\rho_2}\Big{(}\frac{3c_2-2c_1}{3}\Big{)}R^2+c_2R^{1/2}+\Lambda,
\end{equation}
with $\Lambda=\sum_i 
S_i(c_2-2c_1)$. Thus, by looking relations (\ref{newsolutionsnoneulersmallrholimit}) and 
(\ref{frgenerlargetlimitbehvr}), we deduce that the effect of the matter fluids is to 
introduce a cosmological constant plus subdominant curvature corrections. 
This result is quite similar to the reconstructed $F(R)$ which produces the superbounce 
solution given in relation (\ref{newsolutionsnoneulersssslargec}), but now the late-time 
behavior is different. 
However, note that relation (\ref{newsolutionsnoneulersssslargec}) gives the superbounce 
generating $F(R)$ gravity for all values of the Ricci curvature $R$, whilst  
(\ref{newsolutionsnoneulersmallrholimit}) is valid only for small values of the curvature 
(large cosmological times). Thereby, the result (\ref{newsolutionsnoneulersssslargec}) 
has a larger validity region and moreover is phenomenologically more appealing.

\subsubsection{Small-time approximation of LQC ekpyrotic scenario}

In the small-time region the LQC ekpyrotic scale factor and Hubble parameter 
(\ref{scalf}) approximately becomes
\begin{equation}\label{lowtbehav}
a(t)\approx 1+\frac{a_0\rho t^2}{2},{\,}{\,}{\,}H(t)\approx a_0\rho t
\end{equation}
and therefore, using relations (\ref{efoldpoar}) and (\ref{riccinrelat}), the e-fold 
parameter $N$ 
can be expressed as a function of the Ricci scalar $R$ through
\begin{equation}\label{efoldr}
N=\ln \left ( \frac{24a_0\rho+ R}{30a_0\rho}\right ).
\end{equation}
Therefore, the differential equation (\ref{newfrw1}), without taking into account any 
matter fluids, can be cast in the following way
\begin{equation}
\label{ricciscalardiffeqnlarget}
 2\left(72a_0\rho^2-18a_0\rho R-R^2\right)F''(R)+ 3 (4a_0\rho 
+ R) F'(R) 
-5  F(R)=0.
\end{equation} 
This differential equation can be solved using hypergeometric functions, namely
\begin{equation}
F(z)= \,_2F_1(\alpha , \beta , \gamma ; z)
\end{equation} 
where 
\begin{equation}\label{changeofvariable}
z=\frac{6a_0\rho-R}{30a_0\rho}
\end{equation}
with $\lambda_1=6a_0\rho$ and $\lambda_2=-24a_0\rho$, and where $\alpha$, $\beta$ and 
$\gamma$ 
satisfy 
\begin{equation}\label{parafinalnew}
\alpha \beta =\frac{5}{2},{\,}{\,}\alpha +\beta +1=-\frac{3}{2},{\,}{\,}\gamma 
=-\frac{1}{2},
\end{equation}
 which accepts  two sets of solutions, namely 
\begin{eqnarray}
\alpha =\frac{1}{4} \left(-5\pm\sqrt{65}\right),{\,}{\,}{\,}\beta =\frac{1}{4} 
\left(-5\mp\sqrt{65}\right).
\label{albsol}
\end{eqnarray}
Since we are in the large curvature limit, we can expand the above general solution in 
terms of $1/R$, obtaining
\begin{eqnarray}
&&\!\!\!\!\!\!\!\!\!\!\!\!\!\!\!\!  \!\!\!  \!\!  
F(R)\simeq R^{-\alpha } \left(30a_0\rho\right)^\alpha
\left\{
\frac{R\,\Gamma[\beta-\alpha  ]\, \Gamma[\gamma ] 
-\alpha \, \Gamma[\beta -\alpha -1] \,\Gamma[\gamma] 
 } {R\,\Gamma[\beta ]\, \Gamma[\gamma-\alpha]} \right.
\notag 
\\ \ \ \ \ \ \ \ 
&&+36 a_0^2\rho^2 \left\{\alpha  (1+\alpha ) \Gamma[-\alpha +\beta ] \Gamma[\gamma ] 
 \right\}
\notag
\\
&&  \ \ \ \ \ \ \ \ \  \left.\times \frac{\left[(1-\beta +\gamma ) (\gamma-\beta )  -8 
(1+\alpha -\gamma ) (\gamma -\beta )   +16  (1+\alpha -\gamma ) (2+\alpha 
-\gamma )\right]}{2 (1+\alpha -\beta ) (2+\alpha -\beta ) \,\Gamma[\beta ] 
\,\Gamma[\gamma -\alpha] R^2}\right\}\nonumber\\
&&
+\, {\cal{O}}\Big{(}\frac{1}{R}\Big{)}^3.
\label{finalrelowtexpres}
\end{eqnarray}
Hence, given  the values of $\alpha $ and $\beta$ from (\ref{albsol}), we deduce  that 
the most dominant term is the first one, namely
\begin{align}\label{finalrelowtexpres123}
& F(R)\simeq R^{-\alpha } \left(30a_0\rho\right)^\alpha   \left\{
\frac{\Gamma[\beta-\alpha  ]\, \Gamma[\gamma ]
 } {\Gamma[\beta ]\, \Gamma[\gamma-\alpha]} \right\}.
\end{align}

Let us make some comments here. In \cite{oikonomoubounce}  it was shown that the 
large-$R$ reconstructed gravity which generates the LQC ekpyrotic scenario is of a 
different form compared to the above relation. The source of the difference is that in 
the present work we use a more accurate limiting behavior for the small-time scale factor 
(namely relation (\ref{lowtbehav})) comparing to \cite{oikonomoubounce}. In particular, 
apart from the method we used above, there is another  well-known reconstruction, firstly 
developed in  \cite{recontechniques}, where the 
reconstruction is achieved using an auxiliary scalar field.
In reference 
\cite{oikonomoubounce} the auxiliary-field reconstruction was used in order 
to face the same issue, namely to extract $F(R)$ gravities that produce the superbounce 
and LQC ekpyrotic scenario. Concerning the superbounce with ordinary matter fluids 
present, both in the large and small curvature limits, the results are the same to the 
ones  obtained in this work. On the other hand, concerning the LQC ekpyrotic expansion, 
in 
the small-curvature limit 
the two methods give the same results, while in the large-curvature 
limits a difference appears.  The reason behind this difference is traced in 
the 
particular form of the scale factor and Hubble parameter and also their limits when $t$ 
tends to 
zero, a case which corresponds to large curvatures. Specifically, the scale factor of the 
LQC 
ekpyrotic scenario given in relation (\ref{scalf}), when $t$ tends to zero is actually 
equal to one, i.e. $\lim_{t\rightarrow 0}a(t)=1$, and moreover the limit of the Hubble 
parameter (\ref{scalf}) is also zero, namely $\lim_{t\rightarrow 0}H(t)=0$.
This is the result that is taken into account in the auxiliary field reconstruction 
technique and 
is materialized explicitly in reference \cite{oikonomoubounce}, resulting to a different 
$F(R)$ 
gravity in 
the large curvature limit (see also \cite{oikonomoubounce1} for similar results). 

On the other hand, in the approach we adopted in this article, if we consider the 
above limits for the scale factor and Hubble parameter, then the reconstruction 
technique yields trivial results or simply does not work, because $G(N)$ would be 
identically equal to zero. Thus, we need an exact functional dependence of the scale 
factor and of the Hubble parameter, with respect to cosmic time $t$, in order for the 
method to work perfectly. That is why we expanded the scale factor in power series 
of $t$ and we kept the leading order terms as in (\ref{lowtbehav}).
In effect, we could say that in this particular case the auxiliary field method is a bit 
more accurate, to leading order at least.

In summary, the above  artificial difference is traced on the particular form of the 
scale factor and of the Hubble parameter, and if someone was able to solve the 
differential equation without taking any limiting cases (a rather formidable task), then 
there would be no difference between the two approaches. An exemplification to support 
this argument is the superbounce case which we were able to solve analytically, in which 
case we had absolute concordance. Hence, in principle, both reconstruction techniques 
lead to the same results, but extra caution is required if the differential equation that 
yields the reconstructed $F(R)$ gravity, cannot be solved analytically.

Before closing this subsection an important remark is in order. As we demonstrated above,
the loop quantum ekpyrotic cosmology can be described by pure metric $F(R)$ gravity, 
with pure referring to the absence of matter. There exists a very well developed strong 
argument, according to which the full loop quantum evolution is successfully described 
by Palatini $F(R)$ gravity \cite{referee3} and not by metric $F(R)$ theories. Notice 
however that a metric $F(R)$ gravity is mathematically equivalent to Palatini $F(R)$ 
theory in the absence of matter fluids, i.e. for pure vacuum solutions, as those we 
presented in the previous analysis. It would be interesting to study the description of 
the loop quantum ekpyrosis, when it exists, in the presence of matter fields, by using 
both metric and Palatini $F(R)$ gravity. We hope to address this issue in a future 
project.

\section{Superbounce and loop quantum ekpyrotic cosmology from $F(G)$ gravity}
\label{FGrecon}

In this Section we will reconstruct the $F(G)$ form that can lead to the superbounce and 
loop quantum ekpyrotic cosmology realizations. Before analyzing the reconstruction 
procedure, in the following subsection we briefly review  the basic features of 
$F(G)$ gravity and we describe the geometric background we will use in its cosmological 
application. For a detailed account on these issues, see 
\cite{Capozziello:2011et} and references therein.

\subsection{$F(G)$ gravity and cosmology}

 Let us review briefly the $F(G)$ gravitational modification and its cosmological 
application. This class of modified gravity is based on the use of the Gauss-Bonnet 
combination $G=R^2-4R_{\mu \nu}R^{\mu \nu}+R_{\mu \nu \rho 
\sigma}R^{\mu 
\nu \rho \sigma}$, with  $R_{\mu \nu \rho \sigma}$ and $R_{\mu \nu}$ the Riemann and 
Ricci tensors respectively 
\cite{Boulware:1985wk,Wheeler:1985nh,Antoniadis:1993jc,Kanti:1998jd,Nojiri:2005vv}. 
In particular, one can construct gravitational modifications using an arbitrary function 
$F(G)$, which prove to lead to interesting cosmological behavior 
\cite{Nojiri:2005jg,Cognola:2006eg,Davis:2007ta, 
Eynard:2007nq,Jawad:2013wla,newrefs1,newrefs2,fg4,
fg6}.
 
 The $F(G)$ action takes the form \cite{Nojiri:2005jg,Cognola:2006eg}
\begin{equation}\label{actionfggeneral}
\mathcal{S}=\frac{1}{2\kappa^2}\int \mathrm{d}x^4\sqrt{-g}\left [ R+F(G)\right ]+S_m,
\end{equation}
with $\kappa^2=8\pi G$ is the gravitational constant, and  where we have also considered 
the action of the matter sector $S_m$. The corresponding field equations for a general 
metric write as
\begin{eqnarray}
\label{fgr1}
&& \!\!\!\!\!\!\!\!\!\!
R_{\mu \nu}-\frac{1}{2}g_{\mu \nu}F(G)+\left(2RR_{\mu \nu}-4R_{\mu 
\rho}R_{\nu}^{\rho}+2R_{\mu}
^{\rho \sigma \tau}R_{\nu \rho \sigma \tau}-4g^{\alpha \rho}g^{\beta \sigma}R_{\mu \alpha 
\nu \beta}
R_{\rho \sigma}\right)F'(G)\notag 
\\ 
&& \  +4 \left[\nabla_{\rho}\nabla_{\nu}F'(G)\right ] R_{\mu}^{\rho}
-4g_{\mu \nu} \left [\nabla_{\rho}\nabla_{\sigma }F'(G)\right ]R^{\rho \sigma }+4 \left 
[\nabla_{\rho}\nabla_{\sigma }F'(G)\right ]g^{\alpha \rho}g^{\beta \sigma }R_{\mu \alpha 
\nu \beta }
\notag 
\\ 
&& \
-2 \left [\nabla_{\mu}\nabla_{\nu}F'(G)\right ]R+2g_{\mu \nu}\left [\square F'(G) 
\right]R
\notag 
\\
&&\
-4 \left[\square F'(G) \right ]R_{\mu \nu }+4 
\left[\nabla_{\mu}\nabla_{\nu}F'(G)\right]R_{\nu}^{\rho }
=\kappa^2T_{\mu \nu }^m,
\end{eqnarray}
where $T_{\mu \nu}^m$ is the matter energy-momentum tensor arising from $S_m$.
In the case of the FRW metric (\ref{metricformfrwhjkh}), the above equations give rise to 
the two Friedmann equations
\begin{equation}
 \label{eqnsfggrav}
  6H^2+F(G)-GF'(G)+24H^3\dot{G}F''(G)=2\kappa^2\rho_{m}
\end{equation}
\begin{eqnarray}
 &&4\dot{H}+6H^2+F(G)-GF'(G)+16H\dot{G}\left ( \dot{H}+H^2\right ) F''(G)
 \notag\\ 
&&\ \ \ \ \  +8H^2\ddot{G}F''(G)+8H^2\dot{G}^2F'''(G)=-2\kappa^2p_{m},
 \label{eqnsfggrav2}
\end{eqnarray}
where $\rho_m$ and $p_m$ are the matter energy density and pressure respectively. Note 
that in FRW geometry, the Ricci scalar is given by (\ref{ricciscal}), namely 
$R=6(2H^2+\dot{H})$, while $G$ reads 
\begin{equation}\label{gausbonehub}
G=24H^2\left (\dot{H}+H^2 \right ).
\end{equation}

\subsection{Ekpyrotic scenario reconstruction from $F(G)$ gravity}

We now use the reconstruction method developed in Refs. \cite{newrefs1,newrefs2}, for 
finding the 
$F(G)$ gravity, in the large cosmic 
time limit, which corresponds to the late-time era of the ekpyrotic scenario. For an 
alternative 
method to the one we shall employ, see Refs. \cite{fg4,fg6}. We start from the pure $F(G)$ 
gravity 
action,
\begin{equation}\label{actionfggeneralnewrefer}
\mathcal{S}=\int \mathrm{d}x^4\sqrt{-g}\left [ \frac{R}{2\kappa^2}+F(G)\right ],
\end{equation}
and by introducing the auxiliary scalar field $\phi$ the above action becomes
\begin{equation}\label{neweriedsdsssd}
S= \int \mathrm{d}x^4\sqrt{-g}\left [ \frac{R}{2\kappa^2}-V(\phi)-\xi (\phi) G\right]\, .
\end{equation}
Varying the action   (\ref{neweriedsdsssd}) we obtain
\begin{equation}\label{erefaop}
V'(\phi)+\xi'(\phi)G=0\, ,
\end{equation}
and if this equation is solved with respect to $\phi$ we obtain the functional dependence 
of the auxiliary scalar as a function of $G$, namely $\phi=\phi (G)$. Then, by 
substituting into the action 
(\ref{neweriedsdsssd}) we obtain the function $F(G)$, which is
\begin{equation}\label{aop2}
F(G)=-V(\phi (G))-\xi (\phi (G))G\, .
\end{equation}
Hence practically, if one finds $\phi (G)$, the $F(G)$ gravity follows straightforwardly. 
For the flat FRW 
background (\ref{metricformfrwhjkh}) we obtain the following field equations:
\begin{equation}\label{fieldeqn1}
\frac{3}{\kappa^2}H^2-V(\phi)-24H^3\frac{\mathrm{d}\xi}{\mathrm{d}t}=0\, ,
\end{equation}
\begin{equation}\label{fgyudsgufyf}
\frac{1}{\kappa^2}\left(2\dot{H}+3H^2 
\right)-V(\phi)-8H^2\frac{\mathrm{d}^2\xi}{\mathrm{d}t^2}-
16 H \dot{H}\frac{\mathrm{d}\xi}{\mathrm{d}t}-16H^3\frac{\mathrm{d}\xi}{\mathrm{d}t}=0\, 
,
\end{equation}
and upon combining them we obtain
\begin{equation}\label{fgyudsgufyfxfsfs}
\frac{2}{\kappa^2}\dot{H}-8H^2\frac{\mathrm{d}^2\xi}{\mathrm{d}t^2}-16 H 
\dot{H}\frac{\mathrm{d}\xi}{\mathrm{d}t}+8H^3\frac{\mathrm{d}\xi}{\mathrm{d}t}=0\, .
\end{equation}
Solving Eq. (\ref{fgyudsgufyfxfsfs}) with respect to $\xi (t)$ we acquire
\begin{equation}
\label{xi}
\xi (t)=\frac{1}{8}\int^t \mathrm{d}x\frac{\alpha (x)}{H(x)^2}W(x)\, ,
\end{equation}
with $W(t)$ being equal to 
\begin{equation}\label{wayou}
W(t)=\frac{2}{\kappa^2}\int^t \mathrm{d}x\frac{H'(x)}{\alpha (x)} \, ,
\end{equation}
and with the prime denoting differentiation with respect to $x$. 
Combining 
(\ref{fgyudsgufyfxfsfs}) and (\ref{xi}) we obtain
\begin{equation}
\label{vpot}
V(t)=\frac{3}{\kappa^2}H(t)^2-3\alpha (t)H(t)W(t)\, .
\end{equation}

Having the functions $V(t)$ and $\xi(t)$ at hand we can easily obtain the $F(G)$ gravity 
by using Eq. (\ref{aop2}). The large $t$ limit of the scale factor is $\alpha 
(t)=a_0^{\rho /2}t^{\rho }$, and 
the corresponding Hubble rate is $H(t)=\frac{\rho }{t}$, therefore the function $\xi (t)$ 
reads as
\begin{equation}\label{dgfra}
\xi (t)=\frac{t^2}{8 \kappa ^2 \rho  (1+\rho )}\, ,
\end{equation}
and the function $V(t)$ is equal to
\begin{equation}\label{diae}
V(t)=\frac{3 \rho ^2}{t^2 \kappa ^2}-\frac{6 \rho ^2}{t^2 \kappa ^2 (\rho+1)}\, .
\end{equation}
By substituting into Eq. (\ref{erefaop}), and upon solving with respect to $t$, we obtain
\begin{equation}\label{newres}
t=\frac{2^{3/4} 3^{1/4} \left(\rho ^4-\rho ^3\right)^{1/4}}{G^{1/4}}\, ,
\end{equation}
and finally substituting this into Eq. (\ref{aop2}), we obtain the final form of the 
$F(G)$ gravity 
in the large cosmic time limit, namely
\begin{equation}\label{fgfinalformnewapop}
F(G)=-\frac{\sqrt{3/2}  \sqrt{G} \sqrt{(\rho-11) \rho ^3}}{\kappa ^2 \rho  (1+\rho )}\, 
.
\end{equation}
From this relation we can see that the resulting behavior at large $t$ is
$\sim \sqrt{G}$.

\subsection{Superbounce Reconstruction from $F(G)$ Gravity}
 
Let us now reconstruct the $F(G)$ form that leads to the superbounce realization. It 
proves convenient to use the well-known reconstruction technique from
\cite{newrefs1,newrefs2}, 
which 
we briefly described in the previous section. Following the same steps as in the previous 
section, 
we may easily obtain the functions $\xi(t)$ and $V(t)$ of Eqs. (\ref{xi}) and 
(\ref{vpot}). Using 
the Hubble rate and the scale factor from Eqs. (\ref{hubble1}) and (\ref{basicsol1}), we 
easily 
obtain the function $V(t)$ which reads
\begin{equation}\label{gdgf}
V(t)=-\frac{12 \left(c^2-2\right)}{c^4 \left(2+c^2\right) (t_*-t)^2 \kappa ^2}\, ,
\end{equation}
while the function $\xi (t)$ reads,
\begin{equation}\label{djdhxi}
\xi (t)=\frac{c^4 t (t-2 t_*)}{32 \left(2+c^2\right)}\, .
\end{equation}
Having these at hand, we easily solve   (\ref{erefaop}) with respect to $t$, and we 
obtain the 
following solutions:
\begin{equation}\label{solutionsforaopsuperb}
t=t_*-\frac{2\ 2^{3/4} 3^{1/4} \left[\frac{\left(c^2-2\right)   \kappa 
^2G}{c^8}\right]^{1/4}}{\sqrt{G} \kappa },\,\,\, 
t=t_*+\frac{2\ 2^{3/4} 3^{1/4} 
\left[\frac{\left(c^2-2\right) \kappa ^2 G}{c^8}\right]^{1/4}}{\sqrt{G} \kappa }\, .
\end{equation}
Hence, by substituting these solutions into (\ref{aop2}), we easily obtain two 
solutions for the $F(G)$ gravity that generates the superbounce, which read
\begin{equation}\label{gfr2apaop}
F_1(G)=\frac{c^4 \left(t_*^2  \kappa ^2G-16 \sqrt{6} \sqrt{\frac{\left(-2+c^2\right)  
\kappa ^2G}{c^8}}\right)}{32 \left(2+c^2\right) \kappa ^2},\,\,\,F_2(G)=\frac{c^4 
\left(t_*^2  \kappa ^2G-16 \sqrt{6} \sqrt{\frac{\left(-2+c^2\right) \kappa 
^2 G}{c^8}}\right)}{32 \left(2+c^2\right) \kappa ^2}\, .
\end{equation}
We need to note that the method of Refs. \cite{newrefs1,newrefs2} is much more direct in 
comparison 
to the method of Refs. \cite{fg4,fg6}, since in the latter case the procedure is more 
complicated.

\section{Superbounce and loop quantum ekpyrotic cosmology from $F(T)$ gravity}
\label{FTrecon}

In this Section we will repeat the above reconstructions in the case of  $F(T)$ gravity, 
i.e. we desire to reconstruct the $F(T)$ forms that can lead to the 
superbounce and loop quantum ekpyrotic cosmology realizations. We start our analysis by a 
brief introduction to $F(T)$ gravity.

\subsection{$F(T)$ gravity and cosmology}

In general, in the torsional formulations of gravity, one uses as dynamical fields the 
vierbeins  $e^\mu_A$ (we use Greek indices for the coordinate space and capital Latin 
indices for the tangent spacetime), which at each point   of 
spacetime form an orthonormal basis for the tangent space.
The metric tensor is obtained from the
dual vierbein through
\begin{equation}
\label{metricdefin}
g_{\mu\nu}(x)=\eta_{AB}\, e^A_\mu (x)\, e^B_\nu (x),
\end{equation}
with $\eta_{AB}={\rm diag} (1,-1,-1,-1)$. Contrary to General Relativity and its 
curvature-based extensions ($F(R)$, $F(G)$ etc), where one uses the 
torsionless Levi-Civita connection (\ref{christofell}), in the present framework we use 
the curvatureless Weitzenb\"{o}ck connection defined as
$\overset{\mathbf{w}}{\Gamma}^\lambda_{\nu\mu}\equiv e^\lambda_A\:
\partial_\mu
e^A_\nu$ \cite{Weitzenb23}. Hence, all the gravitational information is embedded in the 
torsion tensor, written as 
\begin{equation}
\label{torsiontensor}
{T}^\lambda_{\:\mu\nu}=\overset{\mathbf{w}}{\Gamma}^\lambda_{
\nu\mu}-%
\overset{\mathbf{w}}{\Gamma}^\lambda_{\mu\nu}
=e^\lambda_A\:(\partial_\mu
e^A_\nu-\partial_\nu e^A_\mu).
\end{equation}
Similarly, the contorsion tensor reads as
$K^{\mu\nu}_{\:\:\:\:\rho}\equiv-\frac{1}{2}\Big(T^{\mu\nu}_{
\:\:\:\:\rho}
-T^{\nu\mu}_{\:\:\:\:\rho}-T_{\rho}^{\:\:\:\:\mu\nu}\Big)$, while we additionally define
$
S_\rho^{\:\:\:\mu\nu}\equiv\frac{1}{2}\Big(K^{\mu\nu}_{\:\:\:\:\rho}
+\delta^\mu_\rho
\:T^{\alpha\nu}_{\:\:\:\:\alpha}-\delta^\nu_\rho\:
T^{\alpha\mu}_{\:\:\:\:\alpha}\Big)$.
One can now construct the torsion scalar through contractions of the torsion tensor 
(similarly to the construction of the Ricci scalar from the Riemann tensor) as
  \cite{Hayashi:1979qx,JGPereira,Maluf:2013gaa}
\begin{equation}
\label{torsionsc}
T\equiv\frac{1}{4}
T^{\rho \mu \nu}
T_{\rho \mu \nu}
+\frac{1}{2}T^{\rho \mu \nu }T_{\nu \mu\rho }
-T_{\rho \mu }^{\ \ \rho }T_{\
\ \ \nu }^{\nu \mu }.
\end{equation}

In the simple, teleparallel equivalent of General Relativity,  the
Lagrangian is just $T$. However, inspired by the $F(R)$
extensions, one can extend $T$ to a function $F(T)$, and thus 
the action in the $F(T)$ modified gravity writes as 
\cite{Linder:2010py,Chen:2010va,Dent:2011zz,Bamba:2010wb,Zhang:2011qp,Sharif001,
Capozziello006,Geng:2011aj,Bohmer:2011si,Gonzalez:2011dr,Karami:2012fu,Bamba:2012vg,
Iorio:2012cm,
Rodrigues:2012qua,Capozziello:2012zj,Chattopadhyay:2012eu,Izumi:2012qj,Li:2013xea,
Ong:2013qja,
Otalora:2013tba,
Nashed:2013bfa,Kofinas:2014owa,Harko:2014sja,Hanafy:2014bsa,
Junior:2015bva,Ruggiero:2015oka} 
\begin{eqnarray}
\label{actionFT1}
S = \frac{1}{2 \kappa^2}\int d^4x e \left[F(T)\right]+S_m,
\end{eqnarray}
where $e = \text{det}(e_{\mu}^A) = \sqrt{-g}$, $\kappa^2=8\pi G$ is the gravitational
constant, and we use units where the light speed is equal to 1.  Note that we have also 
added the matter action $S_m$. Therefore, varying the action (\ref{actionFT1}) with 
respect
 to the vierbeins leads to the field equations:
\begin{equation}
\label{FTeoms}
 e^{-1}\partial_{\mu}(ee_A^{\rho}S_{\rho}{}^{\mu\nu})F_T
 +
e_A^{\rho}S_{\rho}{}^{\mu\nu}\partial_{\mu}({T})F_{TT}
-F_Te_{A}^{\lambda}T^{\rho}{}_{\mu\lambda}S_{\rho}{}^{\nu\mu}+\frac{1}{4} e_ { A
} ^ {
\nu
}F(T)  = \frac{\kappa^2}{2}e_{A}^{\rho}\overset {m}T_{\rho}{}^{\nu},
\end{equation}
with $F_{T}=\partial F/\partial T$, $F_{TT}=\partial^{2} F/\partial T^{2}$,
and $\overset{m}{T}_{\rho}{}^{\nu}$  denoting the usual energy-momentum tensor. Note the 
great advantage of $F(T)$ gravity, namely that its field equations are of second order, 
while in $F(R)$ gravity they are of fourth order.

In order to apply $F(T)$ gravity in a cosmological framework, we have to consider an 
ansatz  for the vierbeins that corresponds to the FRW metric (\ref{metricformfrwhjkh}), 
namely 
\begin{equation}
\label{vierbeinsFRW}
e_{\mu}^A={\rm
diag}(1,a,a,a).
\end{equation}
In this case, the field equations (\ref{FTeoms}) give the two Friedmann equations as:
 \begin{eqnarray}
 \label{FTFR1}
&&\frac{TF_T}{3}-\frac{F}{6} +\frac{\kappa^2}{3}\rho_m=0
\\
 \label{FTFR2}
&&\dot{H}(F_T+2T F_{TT})=-\frac{\kappa^2}{2}(\rho_m+p_m),
\end{eqnarray}
   with 
$\rho_m$ and $p_m$ the matter energy density and pressure respectively, and
where
\begin{equation}
\label{TH2}
T=-6H^2,
\end{equation}
as it arises from (\ref{torsionsc}) for the FRW ansatz (\ref{vierbeinsFRW}). Hence, the 
General Relativity equations are obtained for $F(T)=T$.

\subsection{Superbounce Reconstruction from $F(T)$ Gravity}

Let us now suitable reconstruct the $F(T)$ form that can gibe rise to the superbounce 
realization  \cite{bounce4}, i.e. to the scale factor (\ref{basicsol1}), namely
\begin{equation}
\label{basicsol1FT}
a(t)\sim (-t+t_*)^{2/c^2},
\end{equation}
with $t_*$ the big crunch time, and  $c>\sqrt{6}$ the ansatz  parameter. As usual we use 
the e-folding number $N$ through $e^{-N}=\frac{a_0}{a}$, and thus for the Hubble 
parameter we obtain $H=2 a^{-c^2/2}/2$. The great advantage of $F(T)$ reconstruction, 
comparing to $F(R)$ reconstruction of subsection \ref{FRsuperbouncereconstr}, is that the 
torsion scalar $T$ is related straightaway with the Hubble parameter through (\ref{TH2}), 
while the Ricci scalar is related to both the Hubble parameter and its derivative. Hence, 
in the present case we can immediately express $N$ as a function of $T$ as
 \begin{equation}\label{efoldNFT}
N=-\frac{1}{c^2}\ln \left(\frac{T}{\tilde{A}}\right),
\end{equation}
 with 
  \begin{equation}
  \label{tildeA}
\tilde{A}=-\frac{24}{c^4} a_0^{-c^2}.
\end{equation}
Additionally, using the matter  energy density  conservation, we can express $\rho_m$ as 
\begin{equation}\label{mattenrgydens}
\rho_m =\sum_i\rho_{i0}a_0^{-3(1+w_i)}e^{-3N(T)(1+w_i)}.
\end{equation}

Inserting these in the first Friedmann equation  (\ref{FTFR1}) we  obtain
 \begin{align}
\label{bigdiffgeneral1FT}
T\frac{\mathrm{d}F(T)}{\mathrm{d}T}-\frac{F(T)}{2}+\sum_iS_{i}T^{
\frac{3(1+w_i)
}{c^2}}=0,
\end{align}
 with
\begin{equation}\label{gfdgfdgf}
S_i=\frac{\kappa^2\rho_{i0}a_0^{-3(1+w_i)}}{\tilde{A}^{\frac{3(1+w_i)}{c^2}}} .
\end{equation}
 The  solution of (\ref{bigdiffgeneral1FT}) provides the exact $F(T)$ form that 
produces the superbounce evolution.

We can immediately see that  (\ref{bigdiffgeneral1FT}) is much more simple than the 
corresponding equation  (\ref{bigdiffgeneral1}) of the $F(R)$ case. In particular, in the 
case of absence of matter fields, the solution reads
\begin{equation}
\label{FTsol1}
F(T)=c_1 \left(-T\right)^{\frac{1}{2}},
\end{equation}
where $c_1$ is an arbitrary parameter (note that the presence of the minus sign in 
front of $T$ was expected, since $T=-6H^2<0$ in the conventions used in this work). 
Hence, simple teleparallel equivalent of General Relativity, without matter fields, 
cannot 
give rise to the superbounce evolution. However, when matter fields are present, 
the corresponding $F(T)$ is found to be   
\begin{eqnarray}
\label{newsolutionsnoneulerssssFT}
F(T)=c_1 \left(-T\right)^{\frac{1}{2}}- \sum_i Q_i 
 \left(-T\right)^{\frac{3(1+w_i)}{c^2}},
\end{eqnarray}
where 
\begin{eqnarray}
Q_i=\frac{2\kappa^2 c^2 \rho_{i0} a_0^{-3(1+w_i)} }{6(1+w_i)-c^2}.
\end{eqnarray}
As we observe, for $c=3(1+w_i)$  ($c=3$ for simple dust matter) we find that the 
superbounce is realized in the case of standard teleparallel equivalent of General 
Relativity, i.e. standard gravity, plus corrections. Finally, note that the above $F(T)$ 
form is in general different than the one required for matter bounce realization in 
$F(T)$ gravity \cite{Cai:2011tc}.

\subsection{Loop quantum cosmology ekpyrotic scenario reconstruction from $F(T)$ gravity}

Let us now reconstruct the $F(T)$ form that gives rise to the LQC ekpyrotic scenario 
(see also \cite{Amoros:2013nxa} for a different approach). As 
we mentioned in subsection \ref{FRLQCreconstr}, the desirable scale factor and 
Hubble parameter should have the form (\ref{scalf}), namely
\begin{equation}\label{scalf2}
a(t)=\left ( a_0t^2+1 \right )^{\rho/2},{\,}{\,}{\,}H(t)=\frac{2a_0\rho t}{a_0t^2+1},
\end{equation}
with $0<\rho\ll 1$, and where $\rho_c$ is the critical density and $a_0=\frac{8\pi 
G\rho_c }{3\rho^2}$. The Hubble parameter can then be expressed as
\begin{equation}\label{hpscfss2}
H^2=4\rho^2a_0\left (a^{\frac{6}{\rho}}-a^{\frac{4}{\rho}}\right),
\end{equation}
and since in $F(T)$ gravity $T=-6 H^2$ (relation (\ref{TH2})), we can immediately find 
that
\begin{equation}\label{TNFTLQC}
T=-24\rho^2a_0\left (a^{\frac{6}{\rho}}-a^{\frac{4}{\rho}}\right).
\end{equation}
Therefore, introducing the e-folding number $N=\ln(a/a_0)$ and eliminating $a$ we can 
express $N$ as a function of $T$ through
\begin{equation}
N(T)=\ln 
\left\{\frac{3^{\frac{-\rho}{2}}}{a_0}\left[1+\left(\frac{2}{M_2}\right)^{\frac{1}{3}}
+\left(\frac{M_2}{2}\right)^{\frac{1}{3}}
\right ]^{\frac{\rho}{2}} \right\},
\label{efoldrssFT}
\end{equation}
with $M_2=2-\frac{9T}{8\rho^2a_0}+\frac{3\sqrt{3}}{\rho^2 
a_0}\sqrt{\frac{3}{64}T^2-\frac{\rho^2a_0 T}{6}}$.
Finally, inserting $N(T)$ from  (\ref{efoldrssFT}) into $\rho_m$-equation 
(\ref{mattenrgydens}) and then into the Friedmann equation (\ref{FTFR1})  we obtain 
 \begin{equation}
\label{bigdiffgeneral1FTLQC}
T\frac{\mathrm{d}F(T)}{\mathrm{d}T}-\frac{F(T)}{2}+\sum_i X_{i}
  \left[1+\left(\frac{2}{M_2}\right)^{\frac{1}{3}}
+\left(\frac{M_2}{2}\right)^{\frac{1}{3}}
\right ]^{-\frac{3\rho(1+w_i)}{2}}=0,
\end{equation}
with 
\begin{equation}\label{gfdgfdgf}
X_i=3^{\frac{3\rho(1+w_i)}{2}}\kappa^2\rho_{i0}.
\end{equation}
The solution of the above differential equation will give the $F(T)$ form that generates 
the LQC ekpyrotic scenario. Unfortunately, this equation cannot be solved analytically, 
and hence, similarly to subsection  \ref{FRLQCreconstr}, we solve it in the large 
and small time regimes separately.

\subsubsection{Large-time approximation of LQC ekpyrotic scenario}

At large times the required scale factor  (\ref{scalf2}) becomes $a(t)\approx 
a_0t^{\rho}$, and thus $H^2=\rho^2 \left(a_0/a\right)^{2/\rho}$. Thus, the $N(T)$ 
expression is significantly simplified, and it reads 
\begin{equation}
N(T)=\frac{\rho}{2}\ln \left(-\frac{6\rho^2}{T}\right).
\label{efoldrssFT2}
\end{equation}
Hence, inserting $N(T)$ from  (\ref{efoldrssFT2}) into $\rho_m$-equation 
(\ref{mattenrgydens}) and then into the Friedmann equation (\ref{FTFR1})  we obtain 
 \begin{equation}
\label{bigdiffgeneral1FTLQC2}
T\frac{\mathrm{d}F(T)}{\mathrm{d}T}-\frac{F(T)}{2}+\sum_i Y_i 
\left(-T\right)^{\alpha_i}=0,
\end{equation}
with
\begin{eqnarray}
&&Y_i=  \kappa^2  \rho_{i0}  
a_0^{-3(1+w_i)}\left(6\rho^2\right)^{-\frac{3(1+w_i)\rho}{2}}\nonumber\\
&&\alpha_i=\frac{3(1+w_i)\rho}{2}.\label{auxilparameters2}
\end{eqnarray}
Equation (\ref{bigdiffgeneral1FTLQC2})  can be easily solved as:
 \begin{equation}
 \label{LQCFTlargetimes}
F(T)=c_1 \left(-T\right)^{\frac{1}{2}}- \sum_i \frac{2Y_i}{2\alpha_i-1}  
\left(-T\right)^{\alpha_i},
\end{equation}
which is the form of $F(T)$ gravity that can generate a  LQC ekpyrotic scenario at large 
times.

\subsubsection{Small-time approximation of LQC ekpyrotic scenario}

At small times the required scale factor  (\ref{scalf2}) becomes $a(t)\approx 
1+\frac{a_0\rho t^2}{2}$, and thus $H(t)\sim a_0\rho t$. Thus, the 
$N(T)$ 
expression becomes
\begin{equation}
N(T)= \ln \left(\frac{1}{a_0}-\frac{T}{12\rho a_0^2}\right).
\label{efoldrssFT3}
\end{equation}
Therefore, inserting $N(T)$ from  (\ref{efoldrssFT3}) into $\rho_m$-equation 
(\ref{mattenrgydens}) and then into the Friedmann equation (\ref{FTFR1})  we obtain 
 \begin{equation}
\label{bigdiffgeneral1FTLQC3}
T\frac{\mathrm{d}F(T)}{\mathrm{d}T}-\frac{F(T)}{2}+\sum_i Z_i 
\left(1-\frac{T}{12a_0\rho}\right)^{\alpha_i}=0,
\end{equation}
with
\begin{eqnarray}
&&Z_i=  \kappa^2  \rho_{i0}
\nonumber\\
&&\alpha_i=-3(1+w_i).
\label{auxilparameters3}
\end{eqnarray}
Equation (\ref{bigdiffgeneral1FTLQC3})  has the solution
 \begin{equation}
 \label{LQCFTsmalltimes}
F(T)=c_1 \left(-T\right)^{\frac{1}{2}}+2\sum_i  
Z_i\,\,_2F_1\left(-\frac{1}{2},-\alpha_i;\frac{1}{2};\frac{T}{12 \rho a_0}\right),
\end{equation}
with $\,_2F_1(a,b;c;z)$ the hypergeometric function. Hence, this is the form of $F(T)$ 
gravity that can generate a  LQC ekpyrotic scenario at small times. Finally, since at 
early times we are in the large torsion regime ($T\gg1$), we can approximate the above 
solution as
  \begin{equation}
 \label{LQCFTsmalltimes2}
F(T)\approx c_1 \left(-T\right)^{\frac{1}{2}}+2\sum_i  
Z_i    \left\{\frac{\sqrt{\frac{\pi}{12a_0\rho}}\, 
\Gamma[\frac{1}{2}-\alpha_i]}{\Gamma[-\alpha_i]}\left(-T\right)^{\frac{1}{2}}
+\frac{\left(\frac{1}{12a_0\rho}\right)^{\alpha_i} 
}{1-2\alpha_i}\left(-T\right)^{\alpha_i}
\right\} .
\end{equation}

\section{Stability of $F(R)$, $F(G)$ and $F(T)$ superbounce and LQC 
ekpyrosis}
\label{Stability}

In the previous Sections we investigated the superbounce and LQC ekpyrosis in the 
framework of $F(R)$, $F(G)$ and $F(T)$ modified gravities. A crucial issue in every 
cosmological evolution is whether at the perturbation level it exhibits instabilities, 
which thus could constrain or exclude the scenario.  In this Section, we perform such 
stability investigation, by linearly perturbing the above solutions. We mention that we 
are interested in the pure gravity case, in order to obtain a first picture for the 
behavior. The study of the full perturbation behavior, including the matter fields, lies 
beyond the scope of this work, and will be addressed in a separate project.

\subsection{Stability in the $F(R)$ reconstructions}

In the context of the reconstruction method \cite{importantpapers3}, the stability study 
was performed in detail in \cite{sergeistability}, the notation and formalism of which we 
adopt in this Section. The stability of the solutions which we obtained in the previous 
Sections, can be examined by performing linear perturbations of the $F(R)$ solution and 
specifically of the auxiliary function $G(N)$. We consider its perturbation to be of the 
form \cite{sergeistability}
\begin{equation}\label{pert1}
G(N)=g(N)+\delta g(N),
\end{equation}
and we insert it in the first FRW equation  (\ref{newfrw1modfrom}). 
Since the background quantity $g(N)$ satisfies exactly equation (\ref{newfrw1modfrom}), 
we easily acquire the equation for the perturbation $\delta g(N)$, namely
\begin{eqnarray}
&& \!\!\!\!\!\!
g(N)\frac{\mathrm{d}^2F(R)}{\mathrm{d}R^2}\Big{|}_{R=R_1}\delta ''g(N) \notag
\\
&&
+\left\{3g(N)\left 
[4g'(N)+g''(N)\right]\frac{\mathrm{d}^3F(R)}{\mathrm{d}R^3}\Big{|}_{R=R_1}
 +\left [3g(N)-\frac{1}{2}g'(N)\right ]\frac{\mathrm{d}^2F(R)}{\mathrm{d}R^2}\Big{|}_{
R=R_1}\right\}\delta 'g(N) \notag
\\
&&
+\left\{
12g(N)\left[4g'(N)+g''(N)\right]\frac{\mathrm{d}^3F(R)}{\mathrm{d}R^3}\Big{|}_{R=R_1}
\right.
\notag \\ 
&&\left. 
\ \ \ \ 
\ \, +\left[-4g(N)+2g'(N)+g''(N)\right]\frac{\mathrm{d}^2F(R)}{\mathrm{d}R^2}\Big{|}_{ 
R=R_1 } +\frac{1}{3}\frac{\mathrm{d}F(R)}{\mathrm{d}R}\Big{|}_{R=R_1}\right\}
\delta g(N)=0,
\label{stabpert1}
\end{eqnarray}
with $R_1=3g'(N)+12g(N)$. From this equation one may directly extract the stability 
conditions for the perturbations of $G(N)$. In particular, they read
\begin{equation}\label{st0}
J_1=\frac{6[4 g'(N)+g''(N)]F'''(R)}{F''(R)}+6-\frac{g'(N)}{g(N)}>0
\end{equation}
and  
\begin{equation}\label{st01}
J_2=\frac{36 [4 g'(N)+g''(N)] F'''(R)}{F''(R)}-12+\frac{6 g'(N)}{g(N)}+\frac{3 
g''(N)}{g(N)}+\frac{ 
F'(R)}{g(N) F''(R)}>0.
\end{equation}

Let us now examine whether the super bounce and LQC ekpyrotic $F(R)$ forms found in 
Section \ref{FRrecon} satisfy the above stability conditions. Considering the 
superbounce cosmological solution  (\ref{frgenerlargetssss}), without the presence of 
matter field,  the stability conditions (\ref{st0}) and (\ref{st01}) respectively become
\begin{eqnarray}
&&\!\!\!\!\!\!\!\!\!\!\!\!\!\!\!\!\!\!\!
J_1=6+c^2-2 c^2 \left[3^{\rho_1} c_1  Q_1 ^{\rho_1} (\rho_1-2) (\rho_1-1) \rho_1 
 +3^{\rho_2} c_2 Q_1^{\rho_2} 
(\rho_2-2) (\rho_2-1) \rho_2\right]  \notag
\\
&& \ \ \ \ \ \  \ \ \ \ \ 
\times \left[3^{\rho_1} c_1 Q_1^{\rho_1} 
(\rho_1-1) \rho_
1+3^{\rho_2} c_2Q_1^{\rho_2} (\rho_2-1) 
\rho_2\right]>0,
\label{stab1}
\end{eqnarray}
with
$Q_1=-A (c^2-4) e^{-c^2 
N}$,
and
\begin{eqnarray}
&&\!\!\!\!\!\!\!\!\!\!\!\!\!\!\!\!
J_2=-12-6 c^2+3 c^4+e^{c^2 N} \Big{(}c_1Q_2^{\rho_1-1} \rho_1
+c_2 Q_2^{\rho_2-1} 
\rho_2\Big{)}A^{-1} Q_3^{-1}
\notag \\ 
&&\!\!
-12c^2 Q_2
\left[c_1Q_2^{\rho_1-3} (\rho_1-2) (\rho_1-1) \rho_1
  +c_2 Q_2^{\rho_2-3} (\rho_2-2) 
(-1+\rho_
2) \rho_2\right]  Q_3^{-1}>0
\label{st2}
\end{eqnarray}
with $Q_2=3 A e^{-c^2 N}(4 - c^2 )$   and $Q_3=c_1 Q_2^{\rho_1-2} (\rho_1-1) 
\rho_1+c_2 Q_2^{\rho_2-2} (\rho_2-1) \rho_2$.
The most interesting subcase that we found in Section \ref{FRrecon} was when the 
parameter $c$ is large, in which case the above stability conditions become
\begin{eqnarray}\label{st41}
&&J_1= 4 c^2+6>0 
\\
\label{st4}
&&J_2=3 c^4+18 c^2 -12>0.
\end{eqnarray}
Hence, in the interesting case of the large-$c$ regime, the $F(R)$ superbounce is free 
of instabilities.

Considering now the superbounce cosmological solution  (\ref{frgenerlargetssss}), with 
the presence of the matter field, i.e. the $F(R)$ form  (\ref{newsolutionsnoneulerssss}),
the stability conditions (\ref{st0}) and (\ref{st01}) respectively become

\begin{align}
& \!\!\!\!\!\!\!
J_1=6+c^2-6c^2 Q_1 
\Big[3^{\rho_1-3} c_1 
 (\rho_1-2) (\rho_1-1) \rho_1Q_1^{\rho_1-3}+3^{\rho_2-3} c_2 
(\rho_2-2) (\rho_2-1) \rho_2Q_1^{\rho_2-3} 
\notag 
\\ 
&\ \  
+3^{\rho_2-2} 
q_1 
(\rho_2-1) \rho_2 (\rho_2+1)Q_1^{\rho_2-2}
-3^{\delta_i+\rho_2-3} Q_1^{\delta_i+\rho_2-3} q_2 (\delta_i+\rho_2-2) 
(\delta_i+\rho_2-1) 
(\delta_i+\rho_2)
\notag 
\\ 
&\ \ 
+3^{\delta_i+\rho_2-1}  Q_1^{\delta_i+\rho_2-1} 
q_2 (\delta_i+\rho_2) (\delta_i+\rho_2+1) (\delta_i+\rho_2+2)\Big]
\notag 
\\ 
&\ \  
\times\Big[3^{\rho_1-2} c_1  
(\rho_1-1) \rho_1Q_1^{\rho_1-2}
+3^{\rho_2-2} 
c_2
(\rho_2-1) 
\rho_2 Q_1^{\rho_2-2} 
 +3^{\rho_2-1} 
q_1 \rho_2 (\rho_2+1)Q_1^{\rho_2-1} 
\notag 
\\ 
&\ \ \ \ \ \
-3^{\delta_i+\rho_2-2} 
q_2 (\delta_i+\rho_2-1) (\delta_i+\rho_2)Q_1^{\delta_i+\rho_2-2} 
+3^{\delta_i+\rho_2} 
q_2 (\delta_i+\rho_2+1) 
(\delta_i+\rho_2+2)Q_1^{\delta_i+\rho_2} \Big]^{-1}
\label{st5}
\end{align}
with  $Q_1=-A (c^2-4) e^{-c^2 
N}$,
and 
\begin{eqnarray}
&& \!\!\!\!\!\!\!\!\!\!\!\!\!\!\!\!\!\!
J_2=-12-6 c^2+3 c^4+
 A \left(Q_3  +Q_4\right)
\Big\{e^{c^2 N} \Big[c_1 \rho_1 Q_2^{
\rho_1-1} 
 +c_2\rho_2 Q_2^{\rho_2-1} \notag 
\\
&&\ \ \ \     +Q_2^{\rho_2} q_1 
(1+\rho_2)-Q_2^{\delta_i+\rho_2-1} q_2 (\delta_i+\rho_2)
  +Q_2^{\delta_i+\rho_2+1} q_2 
(2+\delta_i+\rho_2)\Big]\Big\}
\nonumber\\
 &&\!\!\!\!\!
+ 36 A c^2 e^{-
c^2 N}   (c^2-4)\left(Q_3  +Q_4\right)^{-1}
\notag 
\\
&&
\times\Big[
c_1 Q_2^{\rho_1-3} 
(\rho_1-2) (\rho_1-1) \rho_1
+c_2 Q_2^{\rho_2-3} (\rho_2-2) 
(\rho_2-1)
 \rho_2
\notag 
\\
&&\ \ \ \ 
+Q_2^{\rho_2-2} q_1 (\rho_2-1) 
\rho_2 (\rho_2+1)
 -Q_2^{\delta_i+\rho_2-3} q_2 
(\delta_
i+\rho_2-2) (\delta_i+\rho_2-1) (\delta_i+\rho_2)
\notag 
\\
&&\ \ \ \ 
+Q_2^{\delta_i+\rho_2-1} q_2 
(\delta_i+\rho_2) (\delta_i+\rho_2+1) (\delta_i+\rho_2+2)\Big],
\label{st6}
\end{eqnarray}
with $Q_2=3 A e^{-c^2 N}(4 - c^2 )$, $Q_3=c_1 Q_2^{\rho_1-2} (\rho_1-1) 
\rho_1+c_2 Q_2^{\rho_2-2} (\rho_2-1) \rho_2$ and
$Q_4= Q_2^{\rho_2-1} q_1 \rho_2 (1+\rho_2)
  -Q_2^{\delta_i+\rho_2-2} q_2 (\delta_
i+\rho_2-1)   (\delta_i+\rho_2) 
  +Q_2^{\delta_i+\rho_2} 
 q_2 (1+\delta_i+\rho_2) (2+\delta_i+\rho_2)$.
Thus, in the large-$c$ regime they become  
\begin{eqnarray}
\label{st7}
&&J_1\approx6c^2
\\
\label{st8}
&& J_2\approx c^4.
\end{eqnarray}
We mention that by large $c$ we practically mean $c\geq 10$ and we also recall that 
$c>\sqrt{6}$ in order a superbounce exists. From relations (\ref{st7}) and (\ref{st8}) it 
is obvious that $J_1>0$ and also $J_2>0$, and therefore the superbounce generating $F(R)$ 
gravity, with or without the presence of matter, is stable under the perturbation 
(\ref{pert1}).

Considering the LQC ekpyrotic $F(R)$ gravity that we reconstructed in subsection 
\ref{FRLQCreconstr}, we mention that we extracted only approximate solutions in the large 
and small time regimes. Although the consistent perturbation analysis should be performed 
in the full expressions, for completeness we examine the stability of the 
approximate $F(R)$ form given in (\ref{finalrelowtexpres123}). In this case the stability 
conditions (\ref{stab1}) and  (\ref{st2}) become
\begin{align}\label{st9}
J_1=1-2 \alpha+\frac{1}{1-e^N}+\frac{8 (2+\alpha )}{5 e^N-4}>0
\end{align}
and  
\begin{align}\label{st10}
J_2=-12+\frac{9 e^N}{e^N-1}-\frac{60 e^N (2+\alpha )  }{ 5e^N  -4 }-\frac{3 e^N   +12 
\left(e^N-1\right)  }{ 
\left(e^N-1\right)(1+\alpha )   }<0.
\end{align}
 As we can see, for small e-folding values $N$ these conditions are not satisfied and 
thus the approximate solution exhibits instabilities. However, the numerical 
investigation of the exact solution is necessary before we conclude on the stability of 
the $F(R)$ LQC ekpyrotic scenario.

\subsection{Stability in the $F(G)$ reconstructions}

We now examine the stability in the $F(G)$ reconstructions of Section \ref{FGrecon}, 
repeating the steps of the previous subsection. The general $F(G)$ stability analysis was 
performed in \cite{newrefs2}, and thus we 
adopt the notation of that work. Inserting the perturbation (\ref{pert1}) into the 
Friedmann equation (\ref{eqnsfggrav}) in the absence of matter fields, we extract the 
following stability conditions:
\begin{equation}\label{stcondgg}
\frac{J_2}{J_1}>0,{\,}{\,}{\,}\frac{J_3}{J_1}>0,
\end{equation}
where $J_1$ stands for
\begin{eqnarray}
\label{st11}
&&\!\!\!\!\!\!\!\!\!\!\!\!\!\!\!\!\!
J_1=288 g(N)^3 F''(G)
\\
&&\!\!\!\!\!\!\!\!\!\!\!\!\!\!\!\!\!
\label{st12}
J_2=432 g(N)^{2 }\left\{(2 g(N)+g'(N)) F''(G)\right.\notag\\
&&\ \ \ \ \ \ \ \  \ \ \ \ \,  \left. +8 g(N) \left[g'(N)^2+g(N) (4 
g'(N)+g''(N))\right] F''
(G)\right\}
\\
\label{st13}
&&\!\!\!\!\!\!\!\!\!\!\!\!\!\!\!\!\!
J_3=6 \Big\{1+24 g(N)\left\{-8 g(N)^2+3 g'(N)^2+6 
g(N) [3 g'(N)+g''(N)]\right\}F''(G)
\notag
\\ && \, \ +24 g(N) [4 g(N)+g'(N)] \left\{g'(N)^2+g(N) [4 
g(N)+g''(N)]\right\}F''(G)\Big\}.
\end{eqnarray}

In the case of the superbounce $F(G)$ gravity, we found two solutions $F_i(G)$, $i=1,2$, 
given in relation (\ref{finalsuperbouncefgs}). Thus, for the function $F_1(G)$ we 
calculate:
\begin{align}\label{st17}
\frac{J_2}{J_1}=\frac{3}{2} (c^2-2) \left(16 A^2 c^2 e^{-2 c^2 N}-1\right)>0
\end{align}
and
\begin{eqnarray}
&&\!\!\!\!\!\!\!\!\!\!\!\!\!\!\!
\frac{J_3}{J_1}=\frac{\left(-\frac{99}{2}c^6+ 108c^4+ 26c^2 -8 \right)}{2-11c^2}   
   +8A e^{-c^2 
N}+\frac{Ac^2 e^{-c^2 N} \left( 11c^6  -46c^4 +  30c^2-4  \right) }{2-11c^2}      
\notag
\\ 
&& \ -   \sqrt{6}(4-c^4)    \sqrt{ \frac{c^2-2}{11c^2-2} } >0,
\label{st18}
\end{eqnarray}
with  $A=\frac{4}{c^4}a_0^{-c^2}$. Hence the $F_1(G)$ solution is conditionally stable 
(note however that in the large-$c$ region the first condition is not satisfied and thus 
instabilities appear).

Similarly, for the case $F_2(G)$ the stability conditions (\ref{stcondgg}) write as
 \begin{align}
\label{st23}
\frac{J_2}{J_1}=\frac{3}{2} \left(c^2-2\right)\left(16 A^2 c^2 e^{-2 c^2 N}-1\right)>0
\end{align}
and
\begin{eqnarray}
&&\!\!\!\!\!\!\!\!\!\!\!\!\!\!
\frac{J_3}{J_1}=
\frac{ 4A^2\left(16-60c^2-190 c^4+207 c^6
\right)}{Q_a}
 \notag 
 \\&&
 +\frac{4A^3 e^{-c^2 N}\left(-32+ 200 c^2  -152 c^4+122 c^6 -68 c^8+11 
c^{10}\right)}{Q_a}
\notag
 \\
 && 
+\frac{ A e^{c^2 N}\left(16-92c^2+30 c^4-46 c^6+11c^8
\right)}{Q_a}
+\frac{ e^{2 c^2 N}\left(-8 +26c^2 +117 c^4-99c^6 \right)}{Q_a}
\notag 
\\ 
&&
+\frac{16 \sqrt{6} A^2 (2+c^2) (4-4c^2+c^4) \sqrt{(2-11c^2)  (c^2-2)
 }}
{c^2Q_a}>0 ,
\label{tor1}
\end{eqnarray}
where  
$Q_a=(2-11c^2)\left[4 A^2 (c^2-2) +e^{2 c^2 N}\right]$. Hence the $F_1(G)$ solution is 
conditionally stable 
(note however that in the large-$c$ region the first condition is not satisfied and thus 
instabilities appear).

\subsection{Stability in the $F(T)$ reconstructions}

Let us now discuss on the stability of the obtained superbounce and LQC ekpyrosis in 
$F(T)$ gravity. As we have mentioned, a crucial difference, and a main advantage, of 
$F(T)$ gravity, comparing to $F(R)$, $F(G)$ and other curvature gravitational 
modifications, is that its equations of motion are of second and not of fourth order. 
This can be immediately seen in the Friedmann equation (\ref{FTFR1}) of $F(T)$ gravity, 
in which only $H$ and not $\dot{H}$ or $\ddot{H}$ appears, and compare it with the 
Friedmann equations of $F(R)$ and $F(G)$ cases, equation (\ref{frwf1}) and  
(\ref{eqnsfggrav}) respectively, where all  $H$, $\dot{H}$ and $\ddot{H}$ are present. 
Hence, in order to study the stability under linear perturbations in $F(T)$ gravity 
solutions, we cannot apply the method we used in the $F(R)$ and $F(G)$ cases, 
which was based on perturbing the auxiliary function $G(N)\equiv H^2(N)$ as in  
(\ref{pert1}), since in this case the results will be trivial. The method used in the 
$F(R)$ and $F(G)$ cases is strictly speaking a method of perturbations of the dynamical 
system formed by the FRW 
equations. In the $F(T)$ case, the stability 
examination requires to perform the complete perturbation analysis, starting 
from perturbing the vierbeins. 
This 
analysis has been performed in detail in 
\cite{Chen:2010va,Dent:2011zz,Izumi:2012qj}, and therefore we do not repeat it here. 
However, using those results, we can easily deduce that the simple power-law $F(T)$ forms 
of the superbounce  ((\ref{FTsol1}) and (\ref{newsolutionsnoneulerssssFT})) and LQC 
ekpyrotic  ((\ref{LQCFTlargetimes}) and (\ref{LQCFTsmalltimes2})) solutions are stable 
(power-law solutions and especially the square root ansatz are very common in $F(T)$ 
gravity 
\cite{Linder:2010py,Chen:2010va,Dent:2011zz,Bamba:2010wb,Zhang:2011qp,Sharif001,
Capozziello006,Geng:2011aj,Bohmer:2011si,Gonzalez:2011dr,Karami:2012fu,Bamba:2012vg,
Iorio:2012cm,
Rodrigues:2012qua,Capozziello:2012zj,Chattopadhyay:2012eu,Izumi:2012qj,Li:2013xea,
Ong:2013qja,
Otalora:2013tba,
Nashed:2013bfa,Kofinas:2014owa,Harko:2014sja,Hanafy:2014bsa,
Junior:2015bva,Ruggiero:2015oka}).

\section{Conclusions}
\label{Conclusions}

In this work we investigated the superbounce and the loop quantum cosmological ekpyrosis, 
in the framework of various modified gravities. Bouncing solutions can be an alternative 
to inflation and thus it is both necessary and interesting to see whether it can be 
obtained naturally, for suitably chosen classes of gravitational modification. We focused 
our investigation in the $F(R)$, $F(G)$ and $F(T)$ modified gravities and in each case we 
reconstructed the forms of modification that give rise to superbounce and LQC ekpyrotic 
evolutions. The reconstruction methods that were used in this work are not the only ones 
that one could follow. Indeed, depending on the modified gravity that is used, one could 
use different techniques, which usually lead to the same exact results, or similar 
results when approximations are used. We will perform a careful comparison and discussion 
of the  various reconstruction methods in a separate publication.
 
Concerning $F(R)$ gravity, we found that the superbounce evolution at the large-curvature 
regime is obtained from $R+\alpha R^2$, while at small curvatures from  
$R+c_1R^{-1/2}+\Lambda$, which agrees with the results obtained in 
\cite{oikonomoubounce} from a different approach. Similarly, the LQC ekpyrosis is 
realized 
from hypergeometric functions that can be approximated by power-law $F(R)$ forms. 
Concerning  $F(G)$ gravity, we found that the superbounce and 
LQC ekpyrosis are obtained from $F(G)$ forms having $G$, $\sqrt{G}$ and $1/\sqrt{G}$ 
factors. Furthermore, in the case of $F(T)$ gravity, we found that the superbounce is 
generated by power-law $F(T)$ forms, while the LQC ekpyrosis from  hypergeometric $F(T)$
functions that can be approximated to power laws. Moreover, performing a linear 
perturbation analysis, we showed that the obtained solutions are conditionally or fully 
stable.

It is expected that bouncing cosmology should occur due to quantum gravity effects. In 
this respect, the important step in its realization is the construction of a realistic 
bouncing cosmology within a specific effective gravity theory. It seems that 
very similar classes of different effective gravities, i.e. power-law, polynomials and 
hypergeometric functions, naturally lead to realization of the proposed bounce universe. 
This indicates that indeed quantum gravity, which may give the origin to specific 
effective $F(R)$, $F(G)$ or $F(T)$ gravity theories, may be responsible for the 
occurrence of bouncing universe.

Finally, we mention that a crucial issue is to analyze in detail the processing of 
perturbations through the bouncing phase in these scenarios, which would lead to extract 
observable signatures than can be confronted with observations and in particular with the 
Planck data. This is the main test that would lead to constraining or excluding the 
scenarios at hand, similarly to the various inflationary models. In addition, a very 
compelling task related to the perturbations analysis is the deviation from 
scale-invariance in the power spectrum, which would result to a blue 
or red tilt in the spectrum of tensor perturbations. These necessary analyses 
lie beyond the scope of the present work, and are left for a future investigation.

\begin{acknowledgments}
The research by SDO has been supported in part by MINECO (Spain), projects
FIS2010-15640 and FIS2013-44881, and by the Russian Ministry of Education and 
Science. The research of ENS is implemented within the framework 
of the Action ``Supporting Postdoctoral Researchers'' of the Operational Program 
``Education and Lifelong
Learning'' (Actions Beneficiary: General Secretariat for Research and Technology), and is 
co-financed by the European Social Fund (ESF) and the Greek State. 
\end{acknowledgments}



\begin{thebibliography}{99}


 


\bibitem{Guth:1980zm}
  A.~H.~Guth,
  {\it{The Inflationary Universe: A Possible Solution to the Horizon and Flatness 
Problems}},
Phys.\ Rev.\ D {\bf 23}, 347 (1981).


\bibitem{Linde:1981mu}
  A.~D.~Linde,
  {\it{A New Inflationary Universe Scenario: A Possible Solution of the Horizon,
  Flatness, Homogeneity, Isotropy and Primordial Monopole Problems}},
Phys.\ Lett.\ B {\bf 108}, 389 (1982).

  
  
\bibitem{Starobinsky:1982ee}
  A.~A.~Starobinsky,
  {\it{Dynamics of Phase Transition in the New Inflationary Universe Scenario and 
Generation of Perturbations}},
Phys.\ Lett.\ B {\bf 117}, 175 (1982).
  
  
\bibitem{Linde:1983gd}
  A.~D.~Linde,
  {\it{Chaotic Inflation}},
Phys.\ Lett.\ B {\bf 129}, 177 (1983).
  
   
   
  
 
 


\bibitem{mukhanov}
V. Mukhanov, 
{\it{Physical foundations of cosmology}},
Cambridge University Press, Campbridge  (2005).

\bibitem{GorbunovRubakov}
D. S. Gorbunov, V. A. Rubakov, 
{\it{Introduction to the theory of the early universe: 
Cosmological perturbations and inflationary theory}},
 World Scientific,  Hackensack, USA (2011).

 
 
  
 
\bibitem{bicep}
  P.~A.~R.~Ade {\it et al.}  [BICEP2 Collaboration],
 {\it{Detection of B-Mode Polarization at Degree Angular Scales by BICEP2}},
Phys.\ Rev.\ Lett.\  {\bf 112}, 241101 (2014)
[\href{http://xxx.lanl.gov/abs/1403.3985}
{{\tt arXiv:1403.3985}}].


 
  
   
\bibitem{planck}
 P.~A.~R.~Ade {\it et al.}  [Planck Collaboration],
 {\it{Planck 2013 results. XVI. Cosmological parameters}},
Astron.\ Astrophys.\ (2014)
[\href{http://xxx.lanl.gov/abs/1303.5076}
{{\tt arXiv:1303.5076}}].



\bibitem{Novello:2008ra}
  M.~Novello and S.~E.~P.~Bergliaffa,
 {\it{Bouncing Cosmologies}},
  Phys.\ Rept.\  {\bf 463}, 127 (2008),
[\href{http://xxx.lanl.gov/abs/0802.1634}
{{\tt arXiv:0802.1634}}].

 \bibitem{Brandenberger:1993ef}
  R.~H.~Brandenberger, V.~F.~Mukhanov and A.~Sornborger,
 {\it{A Cosmological theory without singularities}},
  Phys.\ Rev.\  D {\bf 48}, 1629 (1993),
[\href{http://xxx.lanl.gov/abs/gr-qc/9303001}
{{\tt arXiv:gr-qc/9303001}}].

 

 
\bibitem{bounce4} 
  M.~Koehn, J.~L.~Lehners and B.~A.~Ovrut,
  {\it{A Cosmological Super-Bounce}},
  Phys.\ Rev.\ D {\bf 90}, 025005 (2014)
  [\href{http://xxx.lanl.gov/abs/1310.7577}
{{\tt arXiv:1310.7577}}].

 

\bibitem{bounce5} 
  Y.~F.~Cai, D.~A.~Easson and R.~Brandenberger,
  {\it{Towards a Nonsingular Bouncing Cosmology}},
  JCAP {\bf 1208}, 020 (2012)
    [\href{http://xxx.lanl.gov/abs/1206.2382}
{{\tt arXiv:1206.2382}}].



\bibitem{bounce3}
  J.~Khoury, B.~A.~Ovrut and J.~Stokes,
 {\it{The Worldvolume Action of Kink Solitons in AdS Spacetime}},
  JHEP {\bf 1208}, 015 (2012)
  [\href{http://xxx.lanl.gov/abs/1203.4562}
{{\tt arXiv:1203.4562}}].
 
  \bibitem{bounce3b}
  T.~Biswas, A.~S.~Koshelev, A.~Mazumdar and S.~Y.~Vernov,
   {\it{Stable bounce and inflation in non-local higher derivative cosmology}},
  JCAP {\bf 1208}, 024 (2012)
    [\href{http://xxx.lanl.gov/abs/1206.6374}
{{\tt arXiv:1206.6374}}].
 
   

 

\bibitem{Lehners:2008vx} 
  J.~L.~Lehners,
  {\it{Ekpyrotic and Cyclic Cosmology}},
  Phys.\ Rept.\  {\bf 465}, 223 (2008)
      [\href{http://xxx.lanl.gov/abs/0806.1245}
{{\tt arXiv:0806.1245}}].



 

  
\bibitem{Qiu:2013eoa} 
  T.~Qiu, X.~Gao and E.~N.~Saridakis,
  {\it{Towards anisotropy-free and nonsingular bounce cosmology with scale-invariant 
perturbations}},
  Phys.\ Rev.\ D {\bf 88}, no. 4, 043525 (2013)
       [\href{http://xxx.lanl.gov/abs/1303.2372}
{{\tt arXiv:1303.2372}}].
 
\bibitem{Cai:2014xxa} 
  Y.~F.~Cai, J.~Quintin, E.~N.~Saridakis and E.~Wilson-Ewing,
  {\it{Nonsingular bouncing cosmologies in light of BICEP2}},
  JCAP {\bf 1407}, 033 (2014)
        [\href{http://xxx.lanl.gov/abs/1404.4364}
{{\tt arXiv:1404.4364}}].

 
  
  
\bibitem{Khoury:2001bz} 
  J.~Khoury, B.~A.~Ovrut, N.~Seiberg, P.~J.~Steinhardt and N.~Turok,
  {\it{From big crunch to big bang}},
  Phys.\ Rev.\ D {\bf 65}, 086007 (2002)
          [\href{http://xxx.lanl.gov/abs/hep-th/0108187}
{{\tt arXiv:hep-th/0108187}}].


 

\bibitem{Khoury:2001zk} 
  J.~Khoury, B.~A.~Ovrut, P.~J.~Steinhardt and N.~Turok,
  {\it{Density perturbations in the ekpyrotic scenario}},
  Phys.\ Rev.\ D {\bf 66}, 046005 (2002)
            [\href{http://xxx.lanl.gov/abs/hep-th/0109050}
{{\tt arXiv:hep-th/0109050}}].


 


\bibitem{Khoury:2001wf}
  J.~Khoury, B.~A.~Ovrut, P.~J.~Steinhardt and N.~Turok,
 {\it{The ekpyrotic universe: Colliding branes and the origin of the hot
big bang}},
  Phys.\ Rev.\  D {\bf 64}, 123522
(2001),
[\href{http://xxx.lanl.gov/abs/hep-th/0103239}
{{\tt arXiv:hep-th/0103239}}].


\bibitem{Erickson:2003zm} 
  J.~K.~Erickson, D.~H.~Wesley, P.~J.~Steinhardt and N.~Turok,
  {\it{Kasner and mixmaster behavior in universes with equation of state w >= 1}},
  Phys.\ Rev.\ D {\bf 69}, 063514 (2004)
              [\href{http://xxx.lanl.gov/abs/hep-th/0312009}
{{\tt arXiv:hep-th/0312009}}].


 
 
\bibitem{Lehners:2013cka} 
  J.~L.~Lehners and P.~J.~Steinhardt,
  {\it{Planck 2013 results support the cyclic universe}},
  Phys.\ Rev.\ D {\bf 87}, no. 12, 123533 (2013)
                [\href{http://xxx.lanl.gov/abs/1304.3122}
{{\tt arXiv:1304.3122}}].


 
  
 
 
\bibitem{Veneziano:1991ek}
  G.~Veneziano,
 {\it{Scale Factor Duality For Classical And Quantum Strings}},
  Phys.\ Lett.\  B {\bf 265}, 287 (1991).


\bibitem{Brustein:1997cv}
  R.~Brustein and R.~Madden,
 {\it{A model of graceful exit in string cosmology}},
  Phys.\ Rev.\  D {\bf 57}, 712 (1998),
[\href{http://xxx.lanl.gov/abs/hep-th/9708046}
{{\tt arXiv:hep-th/9708046}}].

\bibitem{Kehagias:1999vr}
  A.~Kehagias and E.~Kiritsis,
 {\it{Mirage cosmology}},
  JHEP {\bf 9911}, 022 (1999),
[\href{http://xxx.lanl.gov/abs/hep-th/9910174}
{{\tt arXiv:hep-th/9910174}}].

 \bibitem{Shtanov:2002mb}
  Y.~Shtanov and V.~Sahni,
 {\it{Bouncing braneworlds}},
  Phys.\ Lett.\  B {\bf 557}, 1 (2003),
[\href{http://xxx.lanl.gov/abs/gr-qc/0208047}
{{\tt arXiv:gr-qc/0208047}}].

\bibitem{Saridakis:2007cf}
  E.~N.~Saridakis,
 {\it{Cyclic Universes from General Collisionless Braneworld Models}},
  Nucl.\ Phys.\ B {\bf 808}, 224 (2009),
[\href{http://xxx.lanl.gov/abs/0710.5269}
{{\tt arXiv:0710.5269}}].


  
  
\bibitem{Creminelli:2007aq} 
  P.~Creminelli and L.~Senatore,
  {\it{A Smooth bouncing cosmology with scale invariant spectrum}},
  JCAP {\bf 0711}, 010 (2007)
     [\href{http://xxx.lanl.gov/abs/hep-th/0702165}
{{\tt arXiv:hep-th/0702165}}].


 
 
\bibitem{Cai:2010zma}
  Y.~-F.~Cai and E.~N.~Saridakis,
 {\it{Cyclic cosmology from Lagrange-multiplier modified gravity}},
  Class.\ Quant.\ Grav.\  {\bf 28}, 035010 (2011),
[\href{http://xxx.lanl.gov/abs/1007.3204}
{{\tt arXiv:1007.3204}}].



\bibitem{HLbounce}
R.~Brandenberger,
 {\it{Matter Bounce in Horava-Lifshitz Cosmology}},
  Phys.\ Rev.\  D {\bf 80}, 043516 (2009),
[\href{http://xxx.lanl.gov/abs/0904.2835}
{{\tt arXiv:0904.2835}}].

\bibitem{Cai:2009in}
  Y.~F.~Cai and E.~N.~Saridakis,
 {\it{Non-singular cosmology in a model of non-relativistic gravity}},
  JCAP {\bf 0910}, 020 (2009),
[\href{http://xxx.lanl.gov/abs/0906.1789}
{{\tt arXiv:0906.1789}}].
  

 

\bibitem{Martin:2003sf}
  J.~Martin and P.~Peter,
 {\it{Parametric amplification of metric fluctuations through a bouncing
  phase}},
  Phys.\ Rev.\  D {\bf 68}, 103517 (2003),
[\href{http://xxx.lanl.gov/abs/hep-th/0307077}
{{\tt arXiv:hep-th/0307077}}].

 
\bibitem{Saridakis:2010mf} 
  E.~N.~Saridakis and S.~V.~Sushkov,
 {\it{Quintessence and phantom cosmology with non-minimal derivative
coupling}},
  Phys.\ Rev.\ D {\bf 81}, 083510 (2010).
[\href{http://xxx.lanl.gov/abs/1002.3478}
{{\tt arXiv:1002.3478}}].
  
 
\bibitem{Cai:2012ag}
  Y.~F.~Cai, C.~Gao and E.~N.~Saridakis,
  {\it{Bounce and cyclic cosmology in extended nonlinear massive gravity}},
  JCAP {\bf 1210} (2012) 048
     [\href{http://xxx.lanl.gov/abs/1207.3786}
{{\tt arXiv:1207.3786}}].
 
 

\bibitem{Bojowald:2001xe} 
  M.~Bojowald,
  {\it{Absence of singularity in loop quantum cosmology}},
  Phys.\ Rev.\ Lett.\  {\bf 86}, 5227 (2001)
                  [\href{http://xxx.lanl.gov/abs/gr-qc/0102069}
{{\tt arXiv:gr-qc/0102069}}].


 

  
\bibitem{Ashtekar:2007tv} 
  A.~Ashtekar,
  {\it{An Introduction to Loop Quantum Gravity Through Cosmology}},
  Nuovo Cim.\ B {\bf 122}, 135 (2007)
   [\href{http://xxx.lanl.gov/abs/gr-qc/0702030}
{{\tt arXiv:gr-qc/0702030}}].

 
  
\bibitem{Ashtekar:2011ni} 
  A.~Ashtekar and P.~Singh,
  {\it{Loop Quantum Cosmology: A Status Report}},
  Class.\ Quant.\ Grav.\  {\bf 28}, 213001 (2011)
     [\href{http://xxx.lanl.gov/abs/1108.0893}
{{\tt arXiv:1108.0893}}].

 
  
  
\bibitem{Bojowald:2008ik} 
  M.~Bojowald,
  {\it{Consistent Loop Quantum Cosmology}},
  Class.\ Quant.\ Grav.\  {\bf 26}, 075020 (2009)
       [\href{http://xxx.lanl.gov/abs/0811.4129}
{{\tt arXiv:0811.4129}}].

 
  
   

\bibitem{Cailleteau:2012fy} 
  T.~Cailleteau, A.~Barrau, J.~Grain and F.~Vidotto,
  {\it{Consistency of holonomy-corrected scalar, vector and tensor perturbations in Loop 
Quantum Cosmology}},
  Phys.\ Rev.\ D {\bf 86}, 087301 (2012)
         [\href{http://xxx.lanl.gov/abs/1206.6736}
{{\tt arXiv:1206.6736}}].

  

  
  
\bibitem{Haro:2014wha} 
  J.~Haro and J.~Amoros,
  {\it{Viability of the matter bounce scenario in Loop Quantum Cosmology for general 
potentials}},
  JCAP(12) 031 (2014)
 [\href{http://xxx.lanl.gov/abs/1406.0369}
{{\tt arXiv:1406.0369}}].

 
  
   
  
  
  
\bibitem{deHaro:2014kxa} 
  J.~de Haro and J.~Amoros,
  {\it{Viability of the matter bounce scenario in Loop Quantum Cosmology from BICEP2 last 
data}},
  JCAP {\bf 1408}, 025 (2014)
   [\href{http://xxx.lanl.gov/abs/1403.6396}
{{\tt arXiv:1403.6396}}].

  
  
   
\bibitem{Amoros:2014tha} 
  J.~Amoros, J.~de Haro and S.~D.~Odintsov,
  {\it{$R+\alpha R^2$ Loop Quantum Cosmology}},
  Phys.\ Rev.\ D {\bf 89}, no. 10, 104010 (2014)
     [\href{http://xxx.lanl.gov/abs/1402.3071}
{{\tt arXiv:1402.3071}}].

   
  
\bibitem{Cai:2014zga} 
  Y.~F.~Cai and E.~Wilson-Ewing,
  {\it{Non-singular bounce scenarios in loop quantum cosmology and the effective field 
description}},
  JCAP {\bf 1403}, 026 (2014)
       [\href{http://xxx.lanl.gov/abs/1402.3009}
{{\tt arXiv:1402.3009}}].

 



\bibitem{WilsonEwing:2012pu} 
  E.~Wilson-Ewing,
  {\it{The Matter Bounce Scenario in Loop Quantum Cosmology}},
  JCAP {\bf 1303}, 026 (2013)
         [\href{http://xxx.lanl.gov/abs/1211.6269}
{{\tt arXiv:1211.6269}}].

 
 

 
 
 
 \bibitem{bouncepiao} 
  Y.~Cai,Y.~T.~Wang and Y.~S.~Piao,
{\it{Pre-inflationary primordial perturbations}},
             [\href{http://xxx.lanl.gov/abs/1501.01730}
{{\tt arXiv:1501.01730}}].



\bibitem{bouncepiao1} 
 Y.~T~Wang and Y.~S.~Piao,
{\it{Parity violation in pre-inflationary bounce}},
Phys.\ Lett.\ B {\bf 741}, 55 (2014)
             [\href{http://xxx.lanl.gov/abs/1409.7153}
{{\tt arXiv:1409.7153}}].
 
 \bibitem{bounceref1} 
  T. Saidov, A. Zhuk,
{\it{Bouncing inflation in nonlinear $R^2+R^4$ gravitational model}},
Phys.\ Rev.\ D {\bf 81}, 124002 (2010)
             [\href{http://xxx.lanl.gov/abs/1002.4138}
{{\tt arXiv:1002.4138}}].

  
  
  \bibitem{bounceref2} 
  M. K. Parikh,
{\it{No open or flat bouncing cosmologies in Einstein gravity}},
             [\href{http://xxx.lanl.gov/abs/1501.04606}
{{\tt arXiv:1501.04606}}]. 
 

\bibitem{kinezosvergados} 
  Y.~K.~E.~Cheung and J.~D.~Vergados,
{\it{Direct dark matter searches - Test of the Big Bounce Cosmology}},
             [\href{http://xxx.lanl.gov/abs/1410.5710}
{{\tt arXiv:1410.5710}}].

 
  
  

\bibitem{oikonomouvergados}
  V.~K.~Oikonomou, J.~D.~Vergados and C.~C.~Moustakidis,
{\it{Direct Detection of Dark Matter-Rates for Various Wimps}},
  Nucl.\ Phys.\ B {\bf 773}, 19 (2007)
               [\href{http://xxx.lanl.gov/abs/hep-ph/0612293}
{{\tt arXiv:hep-ph/0612293}}].

 
  

\bibitem{oikonomouvergados2}
  J.~D.~Vergados, C.~C.~Moustakidis and V.~Oikonomou,
 {\it{Event rates for WIMP detection}},
  AIP Conf.\ Proc.\  {\bf 878}, 138 (2006)
                 [\href{http://xxx.lanl.gov/abs/hep-ph/0610017}
{{\tt arXiv:hep-ph/0610017}}].


  
  
  
  
  
  


\bibitem{Nojiri:2006ri}
  S.~Nojiri and S.~D.~Odintsov,
      {\it{Introduction to modified gravity and gravitational alternative
for dark
  energy}},
  eConf {\bf C0602061}, 06 (2006), Int.\ J.\ Geom.\ Meth.\ Mod.\ Phys.\ 
{\bf 4}, 115 (2007),
[\href{http://xxx.lanl.gov/abs/hep-th/0601213}
{{\tt arXiv:hep-th/0601213}}].


  
\bibitem{Capozziellbook}
S. Capozziello, V. Faraoni, {\it{Beyond Einstein Gravity}}, Springer, Berlin (2010).



\bibitem{Capozziello:2011et}
  S.~Capozziello and M.~De Laurentis,
  {\it{Extended Theories of Gravity}},
  Phys.\ Rept.\  {\bf 509}, 167 (2011),
  [\href{http://xxx.lanl.gov/abs/1108.6266}
{{\tt arXiv:1108.6266}}].

\bibitem{Capozziello:2002rd} 
  S.~Capozziello,
       {\it{Curvature quintessence}},
  Int.\ J.\ Mod.\ Phys.\ D {\bf 11}, 483 (2002)
    [\href{http://xxx.lanl.gov/abs/gr-qc/0201033}
{{\tt arXiv:gr-qc/0201033}}].


 
  
\bibitem{Carroll:2003wy} 
  S.~M.~Carroll, V.~Duvvuri, M.~Trodden and M.~S.~Turner,
  {\it{Is cosmic speed - up due to new gravitational physics?}},
  Phys.\ Rev.\ D {\bf 70}, 043528 (2004)
      [\href{http://xxx.lanl.gov/abs/astro-ph/0306438}
{{\tt arXiv:astro-ph/0306438}}].


 
  
\bibitem{Capozziello:2003gx} 
  S.~Capozziello, V.~F.~Cardone, S.~Carloni and A.~Troisi,
  {\it{Curvature quintessence matched with observational data}},
  Int.\ J.\ Mod.\ Phys.\ D {\bf 12}, 1969 (2003)
           [\href{http://xxx.lanl.gov/abs/astro-ph/0307018}
{{\tt arXiv:astro-ph/0307018}}].


 
  

\bibitem{Sotiriou:2006qn} 
  T.~P.~Sotiriou and S.~Liberati,
  {\it{Metric-affine f(R) theories of gravity}},
  Annals Phys.\  {\bf 322}, 935 (2007)
               [\href{http://xxx.lanl.gov/abs/gr-qc/0604006}
{{\tt arXiv:gr-qc/0604006}}].

  
  
\bibitem{Capozziello:2006dj} 
  S.~Capozziello, S.~Nojiri, S.~D.~Odintsov and A.~Troisi,
  {\it{Cosmological viability of f(R)-gravity as an ideal fluid and its compatibility 
with a 
matter dominated phase}},
  Phys.\ Lett.\ B {\bf 639}, 135 (2006)
  [\href{http://xxx.lanl.gov/abs/astro-ph/0604431}
{{\tt arXiv:astro-ph/0604431}}].

  
   
  
\bibitem{Faraoni:2007yn} 
  V.~Faraoni,
  {\it{de Sitter space and the equivalence between f(R) and scalar-tensor gravity}},
  Phys.\ Rev.\ D {\bf 75}, 067302 (2007)
    [\href{http://xxx.lanl.gov/abs/gr-qc/0703044}
{{\tt arXiv:gr-qc/0703044}}].

 
  
\bibitem{Hu:2007nk} 
  W.~Hu and I.~Sawicki,
  {\it{Models of f(R) Cosmic Acceleration that Evade Solar-System Tests}},
  Phys.\ Rev.\ D {\bf 76}, 064004 (2007)
      [\href{http://xxx.lanl.gov/abs/0705.1158}
{{\tt arXiv:0705.1158}}].

 
  

\bibitem{Appleby:2007vb} 
  S.~A.~Appleby and R.~A.~Battye,
  {\it{Do consistent $F(R)$ models mimic General Relativity plus $\Lambda$?}},
  Phys.\ Lett.\ B {\bf 654}, 7 (2007)
        [\href{http://xxx.lanl.gov/abs/0705.3199}
{{\tt arXiv:0705.3199}}].

 
  
  
\bibitem{Nojiri:2007as} 
  S.~Nojiri and S.~D.~Odintsov,
  {\it{Unifying inflation with LambdaCDM epoch in modified f(R) gravity consistent with 
Solar System tests}},
  Phys.\ Lett.\ B {\bf 657}, 238 (2007)
          [\href{http://xxx.lanl.gov/abs/0707.1941}
{{\tt arXiv:0707.1941}}].

 
  
  
\bibitem{Dunsby:2010wg} 
  P.~K.~S.~Dunsby, E.~Elizalde, R.~Goswami, S.~Odintsov and D.~S.~Gomez,
  {\it{On the LCDM Universe in f(R) gravity}},
  Phys.\ Rev.\ D {\bf 82}, 023519 (2010)
              [\href{http://xxx.lanl.gov/abs/1005.2205}
{{\tt arXiv:1005.2205}}].


 
   
 
\bibitem{importantpapers3} 
  S.~Nojiri, S.~D.~Odintsov and D.~Saez-Gomez,
  {\it{Cosmological reconstruction of realistic modified F(R) gravities}},
  Phys.\ Lett.\ B {\bf 681}, 74 (2009)
                [\href{http://xxx.lanl.gov/abs/0908.1269}
{{\tt arXiv:0908.1269}}].

 
  


\bibitem{recontechniques} 
  S.~Nojiri and S.~D.~Odintsov,
  {\it{Modified f(R) gravity consistent with realistic cosmology: From matter dominated 
epoch to dark energy universe}},
  Phys.\ Rev.\ D {\bf 74}, 086005 (2006)
   [\href{http://xxx.lanl.gov/abs/hep-th/0608008}
{{\tt arXiv:hep-th/0608008}}].

 
  
  
\bibitem{recontechniques1}
  S.~Carloni, R.~Goswami and P.~K.~S.~Dunsby,
  {\it{A new approach to reconstruction methods in $f(R)$ gravity}},
  Class.\ Quant.\ Grav.\  {\bf 29}, 135012 (2012)
     [\href{http://xxx.lanl.gov/abs/1005.1840}
{{\tt arXiv:1005.1840}}].

 
  
  
\bibitem{recontechniques1b} 
  A.~de la Cruz-Dombriz and A.~Dobado,
  {\it{A f(R) gravity without cosmological constant}},
  Phys.\ Rev.\ D {\bf 74}, 087501 (2006)
       [\href{http://xxx.lanl.gov/abs/gr-qc/0607118}
{{\tt arXiv:gr-qc/0607118}}].

 
  
  

\bibitem{recon3}
  E.~Elizalde, E.~O.~Pozdeeva and S.~Y.~Vernov,
  {\it{Reconstruction Procedure in Nonlocal Models}},
  Class.\ Quant.\ Grav.\  {\bf 30}, 035002 (2013)
         [\href{http://xxx.lanl.gov/abs/1209.5957}
{{\tt arXiv:1209.5957}}].

 
 
  




\bibitem{sergeinojirimodel} 
  S.~Nojiri and S.~D.~Odintsov,
  {\it{Modified gravity with negative and positive powers of the curvature: Unification 
of 
the inflation and of the cosmic acceleration}},
  Phys.\ Rev.\ D {\bf 68}, 123512 (2003)
           [\href{http://xxx.lanl.gov/abs/hep-th/0307288}
{{\tt arXiv:hep-th/0307288}}].

  

 
\bibitem{sergeibabmba}
  K.~Bamba, S.~Nojiri and S.~D.~Odintsov,
  {\it{The Universe future in modified gravity theories: Approaching the finite-time 
future 
singularity}},
  JCAP {\bf 0810}, 045 (2008)
             [\href{http://xxx.lanl.gov/abs/0807.2575}
{{\tt arXiv:0807.2575}}].

  

\bibitem{DeFelice:2010aj}
  Claudia de Rham,
  {\it{Massive Gravity}},
  Living Rev.\ Rel.\  {\bf 17}, 7 (2014)
     [\href{http://xxx.lanl.gov/abs/1401.4173}
{{\tt arXiv:1401.4173}}].

  
 

\bibitem{Nojiri:2010wj}
  S.~Nojiri and S.~D.~Odintsov,
  {\it{Unified cosmic history in modified gravity: from F(R) theory to Lorentz
non-invariant models}},
  Phys.\ Rept.\  {\bf 505}, 59 (2011)
       [\href{http://xxx.lanl.gov/abs/1011.0544}
{{\tt arXiv:1011.0544}}].

 
  
  
\bibitem{Boulware:1985wk} 
  D.~G.~Boulware and S.~Deser,
  {\it{String Generated Gravity Models}},
  Phys.\ Rev.\ Lett.\  {\bf 55}, 2656 (1985).

\bibitem{Wheeler:1985nh}
  J.~T.~Wheeler,
  {\it{Symmetric Solutions to the Gauss-Bonnet Extended Einstein Equations}},
  Nucl.\ Phys.\ B {\bf 268}, 737 (1986).
  
  
  
  
 

\bibitem{Antoniadis:1993jc}
  I.~Antoniadis, J.~Rizos and K.~Tamvakis,
  {\it{Singularity - free cosmological solutions of the superstring effective
action}},
  Nucl.\ Phys.\ B {\bf 415}, 497 (1994)
       [\href{http://xxx.lanl.gov/abs/hep-th/9305025}
{{\tt arXiv:hep-th/9305025}}].

 
\bibitem{Kanti:1998jd}
  P.~Kanti, J.~Rizos and K.~Tamvakis,
  {\it{Singularity free cosmological solutions in quadratic gravity}},
  Phys.\ Rev.\ D {\bf 59}, 083512 (1999)
         [\href{http://xxx.lanl.gov/abs/gr-qc/9806085}
{{\tt arXiv:gr-qc/9806085}}].

 
 
 \bibitem{Nojiri:2005vv}
  S.~Nojiri, S.~D.~Odintsov and M.~Sasaki,
  {\it{Gauss-Bonnet dark energy}},
  Phys.\ Rev.\ D {\bf 71}, 123509 (2005)
           [\href{http://xxx.lanl.gov/abs/hep-th/0504052}
{{\tt arXiv:hep-th/0504052}}].

 
   
  
  

\bibitem{Nojiri:2005jg}
  S.~Nojiri and S.~D.~Odintsov,
  {\it{Modified Gauss-Bonnet theory as gravitational alternative for dark
energy}},
  Phys.\ Lett.\ B {\bf 631}, 1 (2005)
         [\href{http://xxx.lanl.gov/abs/hep-th/0508049}
{{\tt arXiv:hep-th/0508049}}].

  

  
\bibitem{Cognola:2006eg} 
  G.~Cognola, E.~Elizalde, S.~Nojiri, S.~D.~Odintsov and S.~Zerbini,
  {\it{Dark energy in modified Gauss-Bonnet gravity: Late-time acceleration and the 
hierarchy problem}},
  Phys.\ Rev.\ D {\bf 73}, 084007 (2006)
           [\href{http://xxx.lanl.gov/abs/hep-th/0601008}
{{\tt arXiv:hep-th/0601008}}].

  
   


\bibitem{Davis:2007ta}
  S.~C.~Davis,
  {\it{Solar system constraints on f(G) dark energy}},
    [\href{http://xxx.lanl.gov/abs/0709.4453}
{{\tt arXiv:0709.4453}}].

 
  
\bibitem{Eynard:2007nq}
  B.~Eynard and N.~Orantin,
  {\it{Topological expansion of mixed correlations in the hermitian 2 Matrix
Model and x-y symmetry of the F(g) invariants}},
    [\href{http://xxx.lanl.gov/abs/0705.0958}
{{\tt arXiv:0705.0958}}].



 
  
\bibitem{Jawad:2013wla}
  A.~Jawad, S.~Chattopadhyay and A.~Pasqua,
  {\it{Reconstruction of f(G) gravity with the new agegraphic dark-energy
model}},
  Eur.\ Phys.\ J.\ Plus {\bf 128}, 88 (2013).
  
  
   \bibitem{newrefs1}
G.~Cognola, E.~Elizalde, S.~Nojiri, S.~Odintsov and S.~Zerbini,
  {\it{String-inspired Gauss-Bonnet gravity reconstructed from the universe expansion 
history and 
yielding the transition from matter dominance to dark energy}},
  Phys.\ Rev.\ D {\bf 75} (2007) 086002
   [\href{http://xxx.lanl.gov/abs/hep-th/0611198}
 {{\tt arXiv:hep-th/0611198}}].
 
 
 


\bibitem{newrefs2}
 S.~Nojiri, S.~D.~Odintsov and M.~Sami,
  {\it{Dark energy cosmology from higher-order, string-inspired gravity and its 
reconstruction}},
  Phys.\ Rev.\ D {\bf 74}, 046004 (2006)
     [\href{http://xxx.lanl.gov/abs/hep-th/0605039}
 {{\tt arXiv:hep-th/0605039}}].
  

 

\bibitem{fg4}
  K.~Bamba, Z.~K.~Guo and N.~Ohta,
  {\it{Accelerating Cosmologies in the Einstein-Gauss-Bonnet Theory with Dilaton}},
  Prog.\ Theor.\ Phys.\  {\bf 118}, 879 (2007)
        [\href{http://xxx.lanl.gov/abs/0707.4334}
{{\tt arXiv:0707.4334}}].

  

\bibitem{fg6} 
  K.~Bamba, S.~D.~Odintsov, L.~Sebastiani and S.~Zerbini,
  {\it{Finite-time future singularities in modified Gauss-Bonnet and F(R,G) gravity and 
singularity avoidance}},
  Eur.\ Phys.\ J.\ C {\bf 67}, 295 (2010)
          [\href{http://xxx.lanl.gov/abs/0911.4390}
{{\tt arXiv:0911.4390}}].


  
 


\bibitem{Linder:2010py}
  E.~V.~Linder,
  {\it{Einstein's Other Gravity and the Acceleration of the
Universe}},
  Phys.\ Rev.\ D \textbf{81}, 127301 (2010)
                [\href{http://xxx.lanl.gov/abs/1005.3039}
{{\tt arXiv:1005.3039}}].

 
  
\bibitem{Chen:2010va}
  S.~H.~Chen, J.~B.~Dent, S.~Dutta and E.~N.~Saridakis,
     {\it{Cosmological perturbations in f(T) gravity}},
  Phys.\ Rev.\  D {\bf 83}, 023508 (2011)
 [\href{http://xxx.lanl.gov/abs/1008.1250}
{{\tt arXiv:1008.1250}}].
  
  
\bibitem{Dent:2011zz}
  J.~B.~Dent, S.~Dutta and E.~N.~Saridakis,
      {\it{f(T) gravity mimicking dynamical dark energy. Background and
perturbation
  analysis}},
  JCAP {\bf 1101}, 009 (2011)
[\href{http://xxx.lanl.gov/abs/1010.2215}
{{\tt arXiv:1010.2215}}].

 
\bibitem{Bamba:2010wb}
  K.~Bamba, C.~Q.~Geng, C.~C.~Lee and L.~W.~Luo,
      {\it{Equation of state for dark energy in $f(T)$ gravity}},
  JCAP {\bf 1101}, 021 (2011)
 [\href{http://xxx.lanl.gov/abs/1011.0508}
 {{\tt arXiv:1011.0508}}].

 
   
 \bibitem{Zhang:2011qp}
   Y.~Zhang, H.~Li, Y.~Gong, Z.~-H.~Zhu,
  {\it{Notes on $f(T)$ Theories}},
  JCAP {\bf 1107}, 015 (2011)
[\href{http://xxx.lanl.gov/abs/1103.0719}
{{\tt arXiv:1103.0719}}].
 
  \bibitem{Sharif001}
  M.~Sharif, S.~Rani,
  {\it{F(T) Models within Bianchi Type I Universe}},
  Mod.\ Phys.\ Lett.\  {\bf A26}, 1657 (2011)
[\href{http://xxx.lanl.gov/abs/1105.6228}
{{\tt arXiv:1105.6228}}].
 
   
 
 \bibitem{Capozziello006}
  S.~Capozziello, V.~F.~Cardone, H.~Farajollahi and A.~Ravanpak,
     {\it{Cosmography in f(T)-gravity}},
  Phys.\ Rev.\  D {\bf 84}, 043527 (2011)
 [\href{http://xxx.lanl.gov/abs/1108.2789}
 {{\tt arXiv:1108.2789}}].
 
  
  
   \bibitem{Geng:2011aj}
  C.~-Q.~Geng, C.~-C.~Lee, E.~N.~Saridakis, Y.~-P.~Wu,
  {\it{'Teleparallel' Dark Energy}},
  Phys.\ Lett.\  {\bf B704}, 384 (2011)
[\href{http://xxx.lanl.gov/abs/1109.1092}
{{\tt arXiv:1109.1092}}].
  
   \bibitem{Bohmer:2011si}
  C.~G.~Bohmer, T.~Harko and F.~S.~N.~Lobo,
  {\it{Wormhole geometries in modified teleparralel gravity and the energy
conditions}},
  Phys.\ Rev.\ D {\bf 85}, 044033 (2012)
   [\href{http://xxx.lanl.gov/abs/1110.5756}
 {{\tt arXiv:1110.5756}}].
 
\bibitem{Gonzalez:2011dr} 
  P.~A.~Gonzalez, E.~N.~Saridakis and Y.~Vasquez,
    {\it{Circularly symmetric solutions in three-dimensional Teleparallel, f(T) and 
Maxwell-f(T) gravity}},
  JHEP {\bf 1207}, 053 (2012)
     [\href{http://xxx.lanl.gov/abs/1110.4024}
 {{\tt arXiv:1110.4024}}].
 
 


 \bibitem{Karami:2012fu}
    K.~Karami and A.~Abdolmaleki,
  {\it{Generalized second law of thermodynamics in f(T)-gravity}},
  JCAP {\bf 1204}, 007 (2012)
     [\href{http://xxx.lanl.gov/abs/1201.2511}
 {{\tt arXiv:1201.2511}}].
 
  
  
  
\bibitem{Bamba:2012vg} 
  K.~Bamba, R.~Myrzakulov, S.~Nojiri and S.~D.~Odintsov,
  {\it{Reconstruction of $f(T)$ gravity: Rip cosmology, finite-time future singularities 
and 
thermodynamics}},
  Phys.\ Rev.\ D {\bf 85}, 104036 (2012)
       [\href{http://xxx.lanl.gov/abs/1202.4057}
 {{\tt arXiv:1202.4057}}].
 
\bibitem{Iorio:2012cm} 
  L.~Iorio and E.~N.~Saridakis,
     {\it{Solar system constraints on f(T) gravity}},
[\href{http://xxx.lanl.gov/abs/1203.5781}
{{\tt arXiv:1203.5781}}].

 


 \bibitem{Rodrigues:2012qua}
  M.~E.~Rodrigues, M.~J.~S.~Houndjo, D.~Saez-Gomez and F.~Rahaman,
  {\it{Anisotropic Universe Models in f(T) Gravity}},
  Phys.\ Rev.\ D {\bf 86}, 104059 (2012)
         [\href{http://xxx.lanl.gov/abs/1209.4859}
 {{\tt arXiv:1209.4859}}].
 
  

\bibitem{Capozziello:2012zj}
  S.~Capozziello, P.~A.~Gonzalez, E.~N.~Saridakis and Y.~Vasquez,
    {\it{Exact charged black-hole solutions in D-dimensional f(T) gravity:
torsion vs curvature analysis}},
  JHEP {\bf 1302} (2013) 039
    [\href{http://xxx.lanl.gov/abs/1210.1098}
{{\tt arXiv:1210.1098}}].

\bibitem{Chattopadhyay:2012eu} 
  S.~Chattopadhyay and A.~Pasqua,
    {\it{Reconstruction of f(T) gravity from the Holographic dark energy}},
  Astrophys.\ Space Sci.\  {\bf 344}, 269 (2013)
      [\href{http://xxx.lanl.gov/abs/1211.2707}
{{\tt arXiv:1211.2707}}].

 
\bibitem{Izumi:2012qj} 
  K.~Izumi and Y.~C.~Ong,
    {\it{Cosmological Perturbation in f(T) Gravity Revisited}},
  JCAP {\bf 1306}, 029 (2013)
        [\href{http://xxx.lanl.gov/abs/1212.5774}
{{\tt arXiv:1212.5774}}].

 
  
  
\bibitem{Li:2013xea}
  J.~T.~Li, C.~C.~Lee and C.~Q.~Geng,
  {\it{Einstein Static Universe in Exponential $f(T)$ Gravity}},
  Eur.\ Phys.\ J.\ C {\bf 73}, 2315 (2013)
   [\href{http://xxx.lanl.gov/abs/1302.2688}
 {{\tt arXiv:1302.2688}}].


\bibitem{Ong:2013qja}
  Y.~C.~Ong, K.~Izumi, J.~M.~Nester and P.~Chen,
    {\it{Problems with Propagation and Time Evolution in f(T) Gravity}},
  Phys.\ Rev.\ D {\bf 88} (2013) 2,  024019
     [\href{http://xxx.lanl.gov/abs/1303.0993}
 {{\tt arXiv:1303.0993}}].

 



\bibitem{Otalora:2013tba}
  G.~Otalora,
   {\it{Scaling attractors in interacting teleparallel dark energy}},
  JCAP {\bf 1307}, 044 (2013)
          [\href{http://xxx.lanl.gov/abs/1305.0474}
{{\tt arXiv:1305.0474}}].



\bibitem{Nashed:2013bfa}
  G.~G.~L.~Nashed,
  {\it{Spherically symmetric charged-dS solution in $f(T)$ gravity theories}},
  Phys.\ Rev.\ D {\bf 88}, no. 10, 104034 (2013)
            [\href{http://xxx.lanl.gov/abs/1311.3131}
{{\tt arXiv:1311.3131}}].



  
   

\bibitem{Kofinas:2014owa}
  G.~Kofinas and E.~N.~Saridakis,
   {\it{Teleparallel equivalent of Gauss-Bonnet gravity and its modifications}},
  Phys.\ Rev.\ D {\bf 90}, no. 8, 084044 (2014)
    [\href{http://xxx.lanl.gov/abs/1404.2249}
{{\tt arXiv:1404.2249}}].

\bibitem{Harko:2014sja} 
  T.~Harko, F.~S.~N.~Lobo, G.~Otalora and E.~N.~Saridakis,
  {\it{Nonminimal torsion-matter coupling extension of f(T) gravity}},
  Phys.\ Rev.\ D {\bf 89}, 124036 (2014)
      [\href{http://xxx.lanl.gov/abs/1404.6212}
{{\tt arXiv:1404.6212}}].

 
\bibitem{Hanafy:2014bsa} 
  W.~E.~Hanafy and G.~L.~Nashed,
  {\it{Vacuum energy $f(T)$ decay: Inflation at the open universe}},
               [\href{http://xxx.lanl.gov/abs/1410.2467}
 {{\tt arXiv:1410.2467}}].
 
  
   
  

\bibitem{Junior:2015bva} 
  E.~L.~B.~Junior, M.~E.~Rodrigues, I.~G.~Salako and M.~J.~S.~Houndjo,
  {\it{Reconstruction and Stability of $\Lambda$CDM Model in $f(T,\mathcal{T})$ Gravity}},
             [\href{http://xxx.lanl.gov/abs/1501.00621}
 {{\tt arXiv:1501.00621}}].
 
  
   
\bibitem{Ruggiero:2015oka} 
  M.~L.~Ruggiero and N.~Radicella,
  {\it{Weak-Field Spherically Symmetric Solutions in $f(T)$ gravity}},
               [\href{http://xxx.lanl.gov/abs/1501.02198}
 {{\tt arXiv:1501.02198}}].
 
 
 
  
 
 \bibitem{oikonomoubounce} 
   V.~K.~Oikonomou,
  {\it{Superbounce and Loop Quantum Cosmology Ekpyrosis from Modified Gravity}},
               [\href{http://xxx.lanl.gov/abs/1412.4343}
 {{\tt arXiv:1412.4343}}].
 
  
   
 
 \bibitem{mbs}
   K.~Bamba, A.~N.~Makarenko, A.~N.~Myagky, S.~Nojiri and S.~D.~Odintsov,
  {\it{Bounce cosmology from $F(R)$ gravity and $F(R)$ bigravity}},
  JCAP {\bf 1401}, 008 (2014)
                 [\href{http://xxx.lanl.gov/abs/1309.3748}
 {{\tt arXiv:1309.3748}}].
 
  
  
  
\bibitem{mbm} 
  S.~D.~Odintsov and V.~K.~Oikonomou,
  {\it{Matter Bounce Loop Quantum Cosmology from $F(R)$ Gravity}},
  Phys.\ Rev.\ D {\bf 90}, 124083 (2014)
                 [\href{http://xxx.lanl.gov/abs/1410.8183}
 {{\tt arXiv:1410.8183}}].
 
  

   
  
  
  
  
\bibitem{sekpd}
  S.~Nojiri, S.~D.~Odintsov and D.~Saez-Gomez,
  {\it{Cyclic, ekpyrotic and little rip universe in modified gravity}},
  AIP Conf.\ Proc.\  {\bf 1458}, 207 (2011)
                   [\href{http://xxx.lanl.gov/abs/1108.0767}
 {{\tt arXiv:1108.0767}}].
 
  
\bibitem{sergeistarobinsky}
  L.~Sebastiani, G.~Cognola, R.~Myrzakulov, S.~D.~Odintsov and S.~Zerbini,
  {\it{Nearly Starobinsky inflation from modified gravity}},
  Phys.\ Rev.\ D {\bf 89}, 023518 (2014)
 [\href{http://xxx.lanl.gov/abs/1311.0744}
 {{\tt arXiv:1311.0744}}].
 
  
   \bibitem{referee1}
  P. Singh, K. Vandersloot, G.V. Vereshchagin,
  {\it{Non-singular bouncing universes in loop quantum cosmology }},
  Phys.\ Rev.\ D {\bf 74}, 043510 (2006) 
 [\href{http://arxiv.org/abs/gr-qc/0606032}
 {{\tt gr-qc/0606032}}].
 
 
 \bibitem{referee2}
  T. Cailleteau, P. Singh, K. Vandersloot,
  {\it{Non-singular ekpyrotic-cyclic model in loop quantum cosmology }},
  Phys.\ Rev.\ D {\bf 80}, 124013 (2009)
 [\href{http://xxx.lanl.gov/abs/0907.5591}
 {{\tt arXiv:0907.5591}}].  
          
     
     
\bibitem{oikonomoubounce1} 
  V.~K.~Oikonomou,
  {\it{Loop Quantum Cosmology Matter Bounce Reconstruction from $F(R)$ Gravity Using an 
Auxiliary Field}},
                 [\href{http://xxx.lanl.gov/abs/1412.8195}
 {{\tt arXiv:1412.8195}}].
 
  \bibitem{referee3}
  G. J. Olmo, P. Singh,
  {\it{Effective Action for Loop Quantum Cosmology a la Palatini}},
  JCAP {\bf 0901} 030 (2009)
 [\href{http://xxx.lanl.gov/abs/0907.5591}
 {{\tt arXiv:0806.2783}}].
 
 


 
 
  


 
 \bibitem{Weitzenb23}
  Weitzenb\"{o}ck R.,
  \emph{Invarianten Theorie},
  Nordhoff, Groningen (1923).


\bibitem{Hayashi:1979qx}
  K.~Hayashi and T.~Shirafuji,
     {\it{New general relativity}},
  Phys.\ Rev.\  D {\bf 19}, 3524 (1979)
  [Addendum-ibid.\  D {\bf 24}, 3312 (1982)].


    \bibitem{JGPereira}
R. Aldrovandi and J. G. Pereira, {\it Teleparallel Gravity: An Introduction},
Springer, Dordrecht (2013).


\bibitem{Maluf:2013gaa}
  J.~W.~Maluf,
  {\it{The teleparallel equivalent of general relativity}},
  Annalen Phys.\  {\bf 525}, 339 (2013),
     [\href{http://xxx.lanl.gov/abs/1303.3897}
{{\tt arXiv:1303.3897}}].
 
 
 
\bibitem{Cai:2011tc}
  Y.~-F.~Cai, S.~-H.~Chen, J.~B.~Dent, S.~Dutta and E.~N.~Saridakis,
 {\it{Matter Bounce Cosmology with the f(T) Gravity}},
  Class.\ Quant.\ Grav.\  {\bf 28}, 215011 (2011),
[\href{http://xxx.lanl.gov/abs/1104.4349}
{{\tt arXiv:1104.4349}}].


    


\bibitem{Amoros:2013nxa} 
  J.~Amoros, J.~de Haro and S.~D.~Odintsov,
  {\it{Bouncing loop quantum cosmology from $F(T)$ gravity}},
  Phys.\ Rev.\ D {\bf 87}, 104037 (2013)
  [\href{http://xxx.lanl.gov/abs/1305.2344}
{{\tt arXiv:1305.2344}}].

   
\bibitem{sergeistability}
  S.~Nojiri and S.~D.~Odintsov,
  {\it{Non-singular modified gravity unifying inflation with late-time acceleration and 
universality of viscous ratio bound in F(R) theory}},
  Prog.\ Theor.\ Phys.\ Suppl.\  {\bf 190}, 155 (2011)
   [\href{http://xxx.lanl.gov/abs/1008.4275}
 {{\tt arXiv:1008.4275}}].
 
   

  
  
  

 



\end{thebibliography}
\end{document}